\documentclass[aps,prb,showpacs,twocolumn,superscriptaddress]{revtex4}
\usepackage{bm,color}
\usepackage{graphicx}
\usepackage{amsmath}

\begin{document}
\title {Theory of spin-phonon coupling in multiferroic Mn perovskites}

\author{Masahito Mochizuki}
\affiliation{Department of Applied Physics, University of Tokyo, Tokyo 113-8656, Japan}
\affiliation{Multiferroics Project, ERATO, Japan Science and Technology Agency (JST) c/o Department of Applied Physics, University of Tokyo, Tokyo 113-8656, Japan}

\author{Nobuo Furukawa}
\affiliation{Multiferroics Project, ERATO, Japan Science and Technology Agency (JST) c/o Department of Applied Physics, University of Tokyo, Tokyo 113-8656, Japan}
\affiliation{Department of Physics, Aoyama Gakuin University, Sagamihara 229-8558, Japan}

\author{Naoto Nagaosa}
\affiliation{Department of Applied Physics, University of Tokyo, Tokyo 113-8656, Japan}
\affiliation{Cross-Correlated Materials Research Group (CMRG) and Correlated Electron Research Group (CERG), RIKEN Advanced Science Institute (ASI), Wako 351-0198, Japan}

\begin{abstract}
Magnetoelectric phase diagrams of the rare-earth ($R$) Mn perovskites $R$MnO$_3$ are theoretically studied by focusing on crucial roles of the symmetric magnetostriction or the Peierls-type spin-phonon coupling through extending our previous work [M. Mochizuki $et$ $al$., Phys. Rev. Lett. {\bf 105}, 037205 (2010)]. We first construct a microscopic classical Heisenberg model for $R$MnO$_3$ including the frustrated spin exchanges, single-ion anisotropy, and Dzyaloshinskii-Moriya interaction. We also incorporate the lattice degree of freedom coupled to the Mn spins via the Peierls-type magnetostriction. By analyzing this model using the replica-exchange Monte-Carlo technique, we reproduce the entire phase diagram of $R$MnO$_3$ in the plane of temperature and magnitude of the orthorhombic lattice distortion. Surprisingly it is found that in the $ab$-plane spiral spin phase, the ($\bm S \cdot \bm S$)-type magnetostriction plays an important role for the ferroelectric order with polarization $\bm P$$\parallel$$\bm a$ whose contribution is comparable to or larger than the contribution from the ($\bm S \times \bm S$)-type magnetostriction, whereas in the $bc$-plane spiral phase, the ferroelectric order with $\bm P$$\parallel$$\bm c$ is purely of ($\bm S \times \bm S$) origin. This explains much larger $\bm P$ in the $ab$-plane spiral phase than the $bc$-plane spiral phase as observed experimentally, and gives a clue how to enhance the magnetoelectric coupling in the spin-spiral-based multiferroics. We also predict a noncollinear deformation of the $E$-type spin structure resulting in the finite ($\bm S \times \bm S$) contribution to the ferroelectric order with $\bm P$$\parallel$$\bm a$, and a wide coexisting regime of the commensurate $E$ and incommensurate spiral states, which resolve several experimental puzzles.
\end{abstract}
\maketitle
\section{Introduction}
\label{Sec:Introduction}
Since the magnetic (electric) induction of electric polarization (magnetization) was proposed theoretically by Dzyaloshinskii in 1959~\cite{Dzyaloshinskii59}, the magnetoelectric coupling in solids has attracted a great deal of interest. Many magnetic materials have been demonstrated to exhibit the magnetoelectric effect~\cite{Astrov60,Freeman75}, but the observed effect was very weak. Recently the interest has been revived by discovery of the magnetically induced ferroelectric order, i.e., multiferroic order in perovskite TbMnO$_3$~\cite{Kimura03a}.

An innovative aspect of this discovery is that in TbMnO$_3$, although the lattice structure retains the inversion symmetry, a nontrivial magnetic order breaks the inversion symmetry and induces the ferroelectric polarization~\cite{Schmid94,Hill00,Fiebig05,Khomskii06,Tokura06a,Eerenstein06,Cheong07,Tokura07}. This is in striking contrast to the usual ferroelectrics whose ferroelectricity originates from the crystal structure with inherent broken inversion symmetry. Therefore the magnetoelectric coupling is very strong in TbMnO$_3$, which leads to a lot of intriguing cross-correlation phenomena such as electromagnon excitations~\cite{Pimenov06a,Kida09,Aguilar09,Mochizuki10a,Mochizuki10d}, magnetic-field control of the ferroelectricity~\cite{Kimura03a,Kimura05,MTokunaga09,Abe07,Murakawa08b,Mochizuki10c}, colossal magnetocapacitance~\cite{Kimura05,Goto04,Kagawa09,Schrettle09} and so on.

TbMnO$_3$ shows successive two magnetic phase transitions with lowering temperature, and below the second transition, the ferroelectric polarization $\bm P$ appears and grows as temperature decreases. The emergence of $\bm P$ in this compound is explained by the antisymmetric magnetostriction associated with the cross-product of spins ($\bm S_i \times \bm S_j$) as described by~\cite{Katsura05,Sergienko06a,Mostovoy06}
\begin{equation}
\bm P_{\rm AS}=A \sum_{<i,j>} \bm e_{i,j} \times (\bm S_i \times \bm S_j).
\label{eqn:invDMmec}
\end{equation}
Here $\bm e_{i,j}$ is the unit vector connecting two spin sites $i$ and $j$, and $A$ is a coupling constant determined by the spin-exchange and spin-orbit interactions. This formula implies that two canted spins $\bm S_i $ and $\bm S_j$ can induce an electric polarization $\bm p_{ij}$ via the spin-orbit coupling. Consequently a transverse spiral spin order as a sequence of the canted spins can generate ferroelectric $\bm P_{\rm AS}$. A neutron-scattering experiment for TbMnO$_3$ confirmed that the Mn spins in its multiferroic phase with $\bm P$$\parallel$$\bm c$ rotate within the $bc$ plane to form a transverse spin spiral propagating along the $b$ axis~\cite{Kenzelmann05}. For the $a$, $b$ and $c$ axes, we adopt the $Pbnm$ setting [see Fig.~\ref{Fig03}(a)]. 

\begin{figure}[tdp]
\includegraphics[scale=1.0]{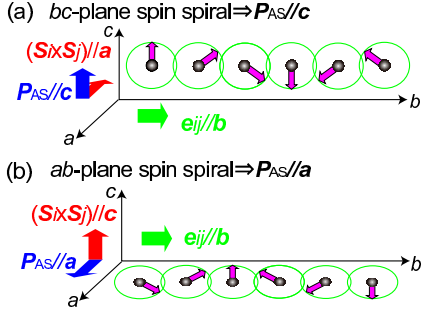}
\caption{(Color online) (a) Relationship between the spiral-plane orientation and the spontaneous ferroelectric polarization $\bm P_{\rm AS}$ in the $bc$-plane spiral spin structure predicted by the spin-current model~\cite{Katsura05,Sergienko06a,Mostovoy06}. (b) That in the $ab$-plane spiral spin structure.}
\label{Fig01}
\end{figure}
This equation also implies that the direction of $\bm P_{\rm AS}$ depends on orientation of the spin spiral plane. In the $bc$-plane ($ab$-plane) spiral spin order, the $\bm P_{\rm AS}$ directs in the $c$ ($a$) direction as shown in Fig.~\ref{Fig01}. This relationship has been confirmed by neutron-scattering experiments~\cite{Kenzelmann05,Yamasaki07a,Arima06,Yamasaki08}. In $R$MnO$_3$ with $R$ being a rare-earth ion, the spiral-plane orientation is determined by a subtle competition between the magnetic anisotropy and the Dzyaloshinskii-Moriya interaction~\cite{Mochizuki09a,Mochizuki09b}. The ground states of TbMnO$_3$ and DyMnO$_3$ exhibit the $bc$-plane spiral order with $\bm P_{\rm AS}$$\parallel$$\bm c$, while the $ab$-plane spiral order with $\bm P_{\rm AS}$$\parallel$$\bm a$ are observed in some solid solutions Eu$_{1-x}$Y$_x$MnO$_3$ and Gd$_{1-x}$Tb$_x$MnO$_3$. 

Another multiferroic phase was theoretically predicted~\cite{Sergienko06b,Picozzi07,Yamauchi08} and experimentally discovered~\cite{Ishiwata10,Ishiwata11,Lorenz07,Pomjakushin09} in $R$MnO$_3$ with much smaller $R$ ions such as Y, Ho, Tm, ..., Lu as well. These compounds exhibit the $E$-type antiferromagnetic ground state where the Mn spins form an up-up-down-down structure. In this state, the symmetric magnetostriction associated with the inner-product of the spins ($\bm S_i \cdot \bm S_j$) induces a ferroelectric polarization $\bm P_{\rm S}$ parallel to the $a$ axis~\cite{Arima06,Kaplan11}, which is given by
\begin{equation}
\bm P_{\rm S} = \sum_{<ij>} \bm \pi_{ij} (\bm S_i \cdot \bm S_j).
\label{eq:Ps}
\end{equation}
Here $\bm \pi_{ij}$ is a form factor which reflects the zigzag MnO chains, and is nonzero because of the absence of inversion symmetry at the center of Mn-O-Mn bond. It is worth mentioning that $\bm P_{\rm S}$ in the $E$-type phase is much larger in magnitude than $\bm P_{\rm AS}$ in the spiral spin phases because the symmetric-magnetostriction mechanism is associated only with the spin-exchange interaction $J$ but not with the spin-orbit interaction or the Dzyaloshinskii-Moriya interaction in contrast to the antisymmetric-magnetostriction mechanism. Typically the magnitude of $\bm P_{\rm S}$ in the $E$-type phase is $\sim$4600 $\mu$C/m$^2$, whereas that of $\bm P_{\rm AS}$ in the spiral spin phases takes $\sim$500 $\mu$C/m$^2$ at most in $R$MnO$_3$.

\begin{figure}[tdp]
\includegraphics[scale=1.0]{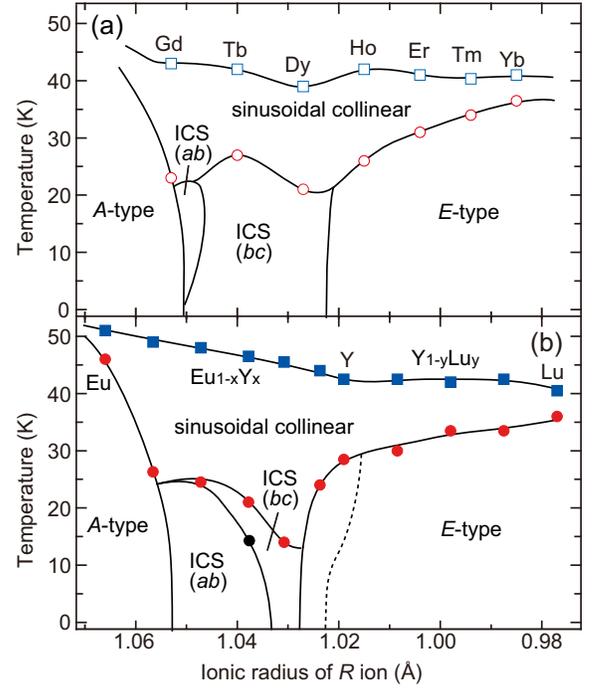}
\caption{(Color online) (a) Experimentally obtained magnetoelectric phase diagram of $R$MnO$_3$ and solid solutions Gd$_{1-x}$Tb$_x$MnO$_3$ and (b) that of solid-solution systems Eu$_{1-x}$Y$_x$MnO$_3$ and Y$_{1-y}$Lu$_y$MnO$_3$ in the plane of temperature and (effective) ionic radius of the $R$ ion~\cite{Ishiwata10}.}
\label{Fig02}
\end{figure}
The perovskite structure of $R$MnO$_3$ is orthorhombically distorted with alternately tilted MnO$_6$ octahedra. Magnitude of this GdFeO$_3$-type distortion varies depending on the size of the $R$ ion. With a smaller $R$ ion, the lattice is more significantly distorted, and the Mn-O-Mn bond angle is reduced more from 180$^{\circ}$. The magnetoelectric phase diagram of $R$MnO$_3$ as a function of the magnitude of the GdFeO$_3$-type distortion or the ionic $R$-site radius ($r_R$) has been studied experimentally~\cite{Ishiwata10,Goto05,Hemberger07,Yamasaki07b,Ivanov06,Ivanov06b}. It has been revealed that following four magnetoelectric phases successively emerge at low temperatures with decreasing $r_R$~\cite{Ishiwata10} [see Figs.~\ref{Fig02}(a) and (b)];
(i) non-ferroelectric $A$-type phase where the Mn spins align ferromagnetically in the $ab$ plane, (ii) ferroelectric $ab$-plane spiral spin phase with $\bm P$$\parallel$$\bm a$, (iii) ferroelectric $bc$-plane spiral spin phase with $\bm P$$\parallel$$\bm c$, and (iv) ferroelectric $E$-type phase with very large $\bm P$$\parallel$$\bm a$. In all these phases, the Mn spins are stacked in a (nearly) staggered manner along the $c$ axis because of the strong interplane antiferromagnetic coupling.

On top of these four phases, there exists a paraelectric sinusoidal collinear spin phase in the intermediate temperature regime where the collinear Mn spins parallel to the $b$ axis are sinusoidally modulated in amplitude. A theoretical model for $R$MnO$_3$ which thoroughly describes severe competitions among these magnetoelectric phases has long been desired.


Here we briefly introduce previous theoretical studies on this issue. In the canonical two-dimensional $J_1$-$J_2$ classical Heisenberg model with ferromagnetic nearest-neighbor interaction $J_1$($<$0) and antiferromagnetic second-neighbor interaction $J_2$($>$0) (see Fig.18 of Ref.~\cite{Mochizuki09b}), the ferromagnetic order is realized for $|J_2/J_1|$$<$0.5, while the spiral order is stabilized for $|J_2/J_1|$$>$0.5. In $R$MnO$_3$, the second-neighbor exchange $J_2$ originates from an indirect overlap of the Mn $3d$ orbitals via the inbetween two O 2$p$ orbitals enhanced by the orthorhombic lattice distortion, and thus is $r_R$ dependent~\cite{Kimura03b}. This can explain the observed phase evolution from the $A$-type phase (i.e., staggered stacking of the ferromagnetic planes) to the spiral phase with decreasing $r_R$. Within this model, however, following experimental observations in $R$MnO$_3$ cannot be reproduced: (i) Stabilities of the specific spin-spiral planes ($ab$- or $bc$-cycloidal spin structures), which depend on $r_R$, temperature, and magnetic field. (ii) Emergence of the $E$-type phase accompanied by ferroelectricity.

In order to reproduce the parameter-dependent spiral-plane directions, magnetic anisotropies are essentially important. Taking into account the orthorhombic lattice structure of $R$MnO$_3$, it is quite natural to examine the single-ion anisotropy and the Dzyaloshinskii-Moriya interaction reflecting the local lattice distortion. By introducing such interactions to the frustrated classical Heisenberg model~\cite{Mochizuki10c,Mochizuki09a,Mochizuki09b}, the authors have successfully reproduced the $r_R$-$T$ and $T$-$H$ phase diagrams of $R$MnO$_3$ with respect to the spin structures. It is noteworthy that the observed 90$^{\circ}$ flop of the spin-spiral plane between $ab$ and $bc$ has been ascribed to the competition between the single-ion anisotropy and the Dzyaloshinskii-Moriya interaction, and thus both ingredients are indispensable.

This model, however, fails to reproduce the $E$-type phase. An attempt to understand the $E$-type phase was performed using an Ising model~\cite{Kimura03b}, which however failed to explain the spiral phases by definition. A two-orbital double-exchange model was examined, and the $E$-type spin order as well as the phase evolution from spiral to $E$-type phases were reproduced~\cite{DongS08}. In the double-exchange model, however, the electron correlation in the $t_{2g}$-orbital sector is neglected although it is usually strong enough to make the system Mott insulating. Thus the mechanism of the $E$-type order in the double-exchange model may not be straightforwardly applicable to the present undoped $R$MnO$_3$ system. Kaplan $et$ $al$ proposed a biquadratic spin interaction originating from the spin-phonon coupling as an origin of the $E$-type order~\cite{Kaplan09,Hayden10}. Bond alternation or staggered modulation of the ferromagnetic exchanges was also proposed for its origin~\cite{Furukawa10}. The latter two works suggest the importance of the spin-lattice coupling to understand the emergence of $E$-type order and its ferroelectricity. 
After these attempts, the entire phase diagram of $R$MnO$_3$ including all the competing magnetoelectric phases was reproduced by a classical Heisenberg model including the Peierls-type spin-lattice coupling~\cite{Mochizuki10b}. This spin-lattice coupling is a source of the ferroelectric polarization associated with the symmetric magnetostriction given by Eq.~(\ref{eq:Ps}).

Here we note that the biquadratic interaction, $-B_{\rm biq} \sum_{<i,j>}(\bm S_i \cdot \bm S_j)^2$, is an effective interaction among spins via the spin-phonon coupling, which is derived by integrating out the phonon degrees of freedom. This interaction favors the collinear spin alignment, and thus stabilizes the $E$-type spin order as compared to the cycloidal orders. However, since this interaction does not contain the phonons or the lattice degrees of freedom explicitly, it is not appropriate to study behaviors of the lattice displacements or the electric polarizations as well as the multiferroic properties. In contrast, the Peierls-type spin-phonon coupling is adequate for such studies.

In this paper, we study theoretically origins and properties of the magnetoelectric phases in $R$MnO$_3$ by focusing on roles of the Peierls-type spin-phonon coupling on the basis of the Monte-Carlo analysis of a spin model. We reveal a large contribution of the ($\bm S \cdot \bm S$)-type magnetostriction to $\bm P$$\parallel$$\bm a$ in the $ab$-plane spiral phase in addition to the ($\bm S \times \bm S$)-type magnetostriction. This finding is quite surprising because the ($\bm S \times \bm S$) mechanism has been considered as a unique origin of the ferroelectric polarization in the spin spiral phase thus far. On the other hand, the $\bm P$$\parallel$$\bm c$ in the $bc$-plane spiral phase is purely of ($\bm S \times \bm S$) origin. This solves a long standing puzzle of much larger $\bm P$ observed in the $ab$-plane spiral phase. This ($\bm S \cdot \bm S$) mechanism can be generally expected in many other spin-spiral-based multiferroics, and gives a clue how to design the enhanced magnetoelectric coupling in materials. We also predict a cycloidal deformation of the $E$-type spin structure, which causes an additional ($\bm S \times \bm S$) contribution to the ferroelectric order with $\bm P$$\parallel$$\bm a$, in addition to the dominant ($\bm S \cdot \bm S$) contribution. In addition, we find a wide regime where the $E$-type and spiral states coexist. On the basis of these findings, we resolve a puzzle in the neutron-scattering experiments for $R$MnO$_3$ with $R$=Y, Ho, and Er. 

The rest of this paper is organized as follows. In Sec.II, we introduce the classical spin model for $R$MnO$_3$ including the Peierls-type spin-phonon coupling. Then we explain the methods for numerical simulations, and physical quantities which we calculate in the simulations in Sec.III. In Sec.IV, we discuss the results for the whole phase diagram, the spiral spin states, the $E$-type state, and the coexistence of the $E$-type and incommensurate states in each subsection. Section V is devoted to the summary. A short report of the present work has been published~\cite{Mochizuki10b}. In addition to the detailed explanation, some further results are presented in this paper.

\begin{figure*}[tdp]
\includegraphics[scale=1.0]{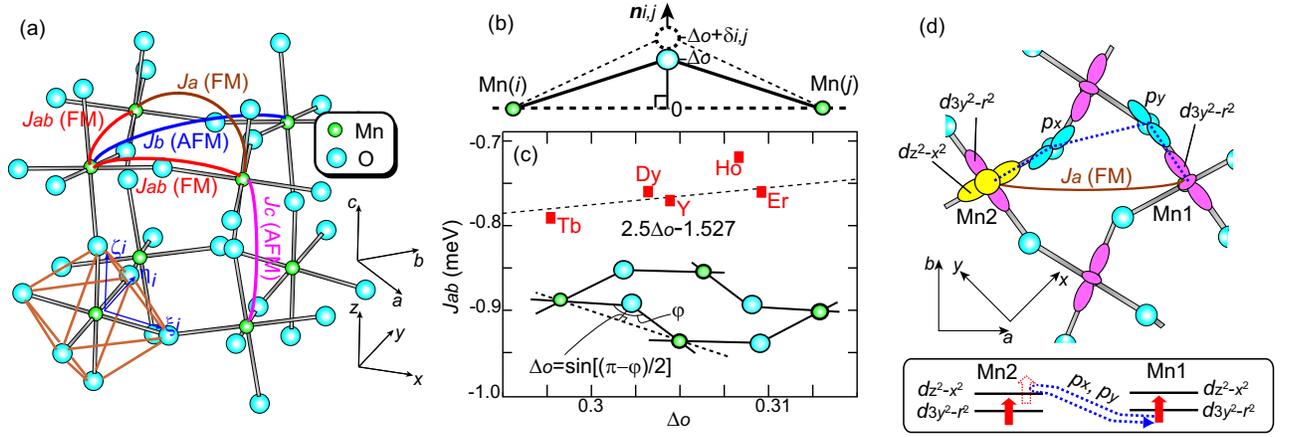}
\caption{(Color online) (a) Spin-exchange interactions in $R$MnO$_3$ and tilted local coordinate axes $\xi_i$, $\eta_i$, and $\zeta_i$ attached to the $i$th MnO$_6$ octahedron. Here FM (AFM) denotes (anti)ferromagnetic exchange interaction. For the spin-exchange interactions, we consider ferromagnetic exchange $J_{ab}$ on the Mn-Mn bonds along the pseudocubic $x$ and $y$ axes, (anti)ferromagnetic exchange $J_a$ ($J_b$) on the in-plane diagonal Mn-Mn bonds along the $a$ ($b$) axis, and antiferromagnetic exchange $J_c$ along the $c$ axis. (b) Mn($i$)-O-Mn($j$) bond in the orthorhombic lattice and local vector $\bm n_{i,j}$. The O ion is displaced from its cubic position (0) to the orthorhombic position ($\Delta_o$) at higher temperatures. At low temperatures, a further shift $\delta_{i,j}$ along $\bm n_{i,j}$ can be induced by the spin-lattice coupling in the presence of magnetic order. (c) $\Delta_o$ vs $J_{ab}$ for several $R$MnO$_3$ compounds calculated in Ref.~\cite{Mochizuki09b}, which gives $J_{ab}'$=$\partial J_{ab}$/$\partial \Delta_o$=2.5. Here $\Delta_o$ is normalized by the MnO bond length. (d) Main exchange path for the next-neighbor ferromagnetic exchange $J_a$ along the $a$ axis [(blue) dotted line].}
\label{Fig03}
\end{figure*}
\section{Model}
\label{Sec:Model}
To describe the Mn 3$d$-spin system in $R$MnO$_3$, we employ a classical Heisenberg model on a cubic lattice~\cite{Mochizuki10b}, in which the Mn $S$=2 spins are treated as classical vectors,
${\bm S}_i$=($\sqrt{S^2-S_{ci}^2}\cos\theta_i$, 
$\sqrt{S^2-S_{ci}^2}\sin\theta_i$, $S_{ci}$) 
with respect to the $a$, $b$, and $c$ axes. The Hamiltonian is given by
\begin{equation}
\mathcal{H}=\mathcal{H}_{\rm ex}+\mathcal{H}_{\rm sia}^D
+\mathcal{H}_{\rm sia}^E+\mathcal{H}_{\rm DM}
+\mathcal{H}_K,
\label{eq:model1}
\end{equation}
with
\begin{eqnarray}
\mathcal{H}_{\rm ex}&=&
\sum_{<i,j>} J_{ij} \bm S_i \cdot \bm S_j,\\
\mathcal{H}_{\rm sia}^D&=&D\sum_{i}S_{\zeta i}^2,\\
\mathcal{H}_{\rm sia}^E&=&
E\sum_{i}(-1)^{i_x+i_y}(S_{\xi i}^2-S_{\eta i}^2),\\
\mathcal{H}_{\rm DM}&=&
\sum_{<i,j>}\bm d_{i,j}\cdot(\bm S_i \times \bm S_j),\\
\mathcal{H}_K&=&K \sum_i(\delta_{i,i+\hat x}^2+\delta_{i,i+\hat y}^2),
\end{eqnarray}
where $i_x$, $i_y$ and $i_z$ represent the integer coordinates of the $i$th Mn ion with respect to the pseudocubic $x$, $y$ and $z$ axes [see Fig.~\ref{Fig03}(a)].

The first term $\mathcal{H}_{\rm ex}$ describes the spin-exchange interactions as shown in Fig.~\ref{Fig03}(a). Since the strength of the nearest-neighbor ferromagnetic coupling in $R$MnO$_3$ sensitively depends on the Mn-O-Mn bond angle, we consider the Peierls-type spin-phonon coupling 
\begin{equation}
J_{ij}=J_{ab}+J_{ab}'\delta_{i,j},
\end{equation}
for the in-plane Mn-O-Mn bonds where $J_{ab}'$=$\partial J_{ab}$/$\partial \delta$. Here $\delta_{i,j}$ and $\delta$ denote a shift of the O ion between $i$th and $j$th Mn ions normalized by the averaged MnO bond length. Note that the O ion in the orthorhombic lattice is already displaced from its cubic position. We consider $\delta_{i,j}$ as a further shift of the position in the presence of magnetic order at low temperatures with respect to the orthorhombic position at higher temperatures. We assume that the shift of the O ion $\delta_{i,j}$ occurs along the local axis $\bm n_{i,j}$ directing from its cubic position (0) to the orthorhombic position ($\Delta_o$) at higher temperature as shown in Fig.~\ref{Fig03}(b). Then the positive (negative) shift decreases (increases) the Mn-O-Mn bond angle. The nearest neighbor ferromagnetic exchange becomes stronger (weaker) as the Mn-O-Mn bond angle increases (decreases), which implies positive $J_{ab}'$.

\begin{table}
\caption{Structural parameters of DyMnO$_3$ from Ref.~\cite{Alonso00}.}
\begin{tabular}{cccccccc}
\hline
 $a$ ($\AA$) & $b$ ($\AA$) & $c$ ($\AA$) & $x_{\rm O_1}$ & $y_{\rm O_1}$ & 
 $x_{\rm O_2}$ & $y_{\rm O_2}$ & $z_{\rm O_2}$ \\
\hline
5.2785 & 5.8337 & 7.3778 & 0.1092 & 0.4642 & 0.7028 & 0.3276&  0.0521\\
\hline
\end{tabular}
\label{tabl:SDEMO}
\end{table}
The second and the third terms, $\mathcal{H}_{\rm sia}^D$ and 
$\mathcal{H}_{\rm sia}^E$, stand for the single-ion anisotropies. 
Here $\xi_i$, $\eta_i$ and $\zeta_i$ are tilted local axes attached to 
the $i$th MnO$_6$ octahedron as shown Fig.~\ref{Fig03}(a). 
The former term makes the magnetization along the $c$ axis hard, 
while the latter term causes alternation of the local easy and hard 
magnetization axes along the $\xi_i$ and $\eta_i$ axes in the $ab$ plane 
owing to the staggered $d_{3x^2-r^2}$/$d_{3y^2-r^2}$ type orbital ordering.
The directional vectors $\bm \xi_i$, $\bm \eta_i$ and $\bm \zeta_i$ with 
respect to the $a$, $b$ and $c$ axes are given by
\begin{eqnarray}
\label{eq:tilax1}
\bm {\xi}_i&=&\left[
\begin{array}{c}
a[0.25+(-1)^{i_x+i_y}(0.75-x_{{\rm O}_2})] \\
b[0.25-(-1)^{i_x+i_y}(y_{{\rm O}_2}-0.25)] \\
c(-1)^{i_x+i_y+i_z}z_{{\rm O}_2} \\
\end{array}
\right], \\
\label{eq:tilax2}
\bm {\eta}_i&=&\left[
\begin{array}{c}
a[-0.25+(-1)^{i_x+i_y}(0.75-x_{{\rm O}_2})] \\
b[0.25+(-1)^{i_x+i_y}(y_{{\rm O}_2}-0.25)] \\
-c(-1)^{i_x+i_y+i_z}z_{{\rm O}_2} \\
\end{array}
\right], \\
\label{eq:tilax3}
\bm {\zeta}_i&=&\left[
\begin{array}{c}
-a(-1)^{i_x+i_y+i_z}x_{{\rm O}_1} \\
b(-1)^{i_z}(0.5-y_{{\rm O}_1}) \\
0.25c \\
\end{array}
\right].
\end{eqnarray}
Here $x_{{\rm O}_2}$, $y_{{\rm O}_2}$ and $z_{{\rm O}_2}$ ($x_{{\rm O}_1}$ and $y_{{\rm O}_1}$) are the coordination parameters of the in-plane (out-of-plane) oxygens, and $a$, $b$ and $c$ are the lattice parameters. For values of these parameters, we use the experimental data of DyMnO$_3$~\cite{Alonso00} throughout the calculations (see Table~\ref{tabl:SDEMO}).

\begin{figure}[tdp]
\includegraphics[scale=1.0]{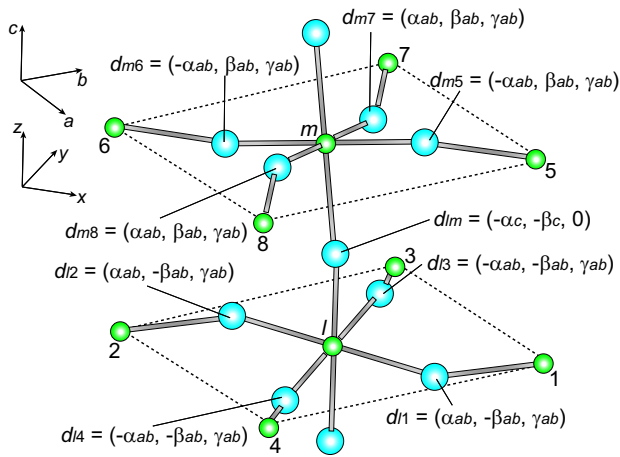}
\caption{(Color online) Dzyaloshinskii-Moriya vectors $\bm d_{i,j}$ associated with the Mn($i$)-O-Mn($j$) bonds.}
\label{Fig04}
\end{figure}
The fourth term, $\mathcal{H}_{\rm DM}$, denotes the Dzyaloshinskii-Moriya interaction~\cite{Dzyaloshinsky58,Moriya60a,Moriya60b}. The vectors $\bm d_{i,j}$ are defined on the Mn($i$)-O-Mn($j$) bonds. Because of the crystal symmetry, they are expressed using five parameters, $\alpha_{ab}$, $\beta_{ab}$, $\gamma_{ab}$, $\alpha_c$, and $\beta_c$, as given in Ref.~\cite{Solovyev96}. Their expressions are given by (see also Fig.~\ref{Fig04}),
\begin{eqnarray}
\bm d_{i,i+\hat x}&=&\left[
\begin{array}{c}
-(-1)^{i_x+i_y+i_z}\alpha_{ab} \\
(-1)^{i_x+i_y+i_z}\beta_{ab} \\
(-1)^{i_x+i_y}\gamma_{ab} \\
\end{array}
\right], \\
\bm d_{i,i+\hat y}&=&\left[
\begin{array}{c}
(-1)^{i_x+i_y+i_z}\alpha_{ab} \\
(-1)^{i_x+i_y+i_z}\beta_{ab} \\
(-1)^{i_x+i_y}\gamma_{ab} \\
\end{array}
\right], \\
\bm d_{i,i+\hat z}&=&\left[
\begin{array}{c}
(-1)^{i_z}\alpha_c \\
(-1)^{i_x+i_y+i_z}\beta_c \\
0 \\
\end{array}
\right].
\label{eq:DMVECS}
\end{eqnarray}
The last term represents the lattice elastic term with $K$ being the elastic constant.

\begin{table}[tdp]
\caption{Model parameters used in the calculations. The energy unit is meV.}
\begin{tabular}{c|ccccc}
\hline
\hline
$\mathcal{H}_{\rm ex}$ & 
$J_{ab}=-0.8$, & $J_a=-0.1$, & $J_c$=1.25 & $J_{ab}'=2.5$ & \\
$\mathcal{H}_{\rm sia}^D$, $\mathcal{H}_{\rm sia}^E$ &
$D$=0.2,      & $E$=0.25   & & & \\
$\mathcal{H}_{\rm DM}$  &
$\alpha_{ab}$=0.1, & $\beta_{ab}$=0.1, & $\gamma_{ab}$=0.14 & & \\
& 
$\alpha_c$=0.42, & $\beta_c$=0.1 &  & & \\
$\mathcal{H}_K$ & 
$K$=500    &            & & &\\
\hline
\hline
\end{tabular}
\label{tabl:MDLPRMS}
\end{table}
The values of $J_{ab}$, $J_c$, $J_b$, $D$, $E$, and five Dzyaloshinskii-Moriya parameters have been microscopically determined or have been estimated in Ref.~\cite{Mochizuki09b} for several $R$MnO$_3$ compounds. Except for $J_b$, they are nearly invariant upon the $R$-site variation in the vicinity of the multiferroic phases. 
We also find that very weak ferromagnetic exchange $J_a$ along the $a$ axis is necessary to produce the $E$-type phase, and adopt $J_a$=$-$0.1 meV. The value of $K$ is chosen so as to reproduce the experimental value of $P$ ($\sim$4600 $\mu$C/m$^2$) in the $E$-type phase, which mostly comes from the ($\bm S \cdot \bm S$) contribution. The model parameters used in the calculations are summarized in Table~\ref{tabl:MDLPRMS}. We obtain the value of $J_{ab}'$ from the $\Delta_o$ dependence of $J_{ab}$ for several $R$ species [see Fig.~\ref{Fig03}(c)], which gives $J_{ab}'$=$\partial J_{ab}$/$\partial \Delta_o$=2.5 meV.

The main exchange path for the ferromagnetic $J_a$ is shown in Fig.~\ref{Fig03}(d), which contains two O $2p$ orbitals. The electron starting from the Mn1 $d_{3y^2-r^2}$ orbital dominantly reaches the unoccupied Mn2 $d_{z^2-x^2}$ orbital via the 2$p_x$ and 2$p_y$ orbitals. This results in the doubly occupied Mn2 ion with $d_{3y^2-r^2}^1 d_{z^2-x^2}^1$ electron configuration as an intermediate state of the perturbation process with respect to the $d$-$p$ and $p$-$p$ transfer integrals. This process favors the parallel spin configuration relative to the antiparallel one since the Hund's-rule coupling reduces energy of the intermediate state, which leads to the ferromagnetic exchange $J_a$.

We treat the antiferromagnetic exchange $J_b$ as a variable which increases (decreases) as $r_R$ decreases (increases). This is because the exchange path for $J_b$ contains two O $2p$ orbitals between two Mn $e_g$ orbitals neighboring along the $b$ direction, and the orthorhombic distortion, whose magnitude is controlled by $r_R$, enhances their $p$-$p$ hybridization~\cite{Kimura03b,Picozzi06}. On the other hand, previous microscopic evaluations of the model parameters showed that the parameters except for $J_b$ are almost insensitive to the $R$-site species in/near the multiferroic phases~\cite{Mochizuki09b}. We find that overall features of the phase evolution upon the $R$-site variation can be reproduced as a function of $J_b$ even without considering the slight $R$-dependence of other parameters.

\section{Method}
\label{Sec:Method}
We investigate finite-temperature properties of the Hamiltonian (\ref{eq:model1}) by using the Monte-Carlo technique. The spin system in $R$MnO$_3$ has frustrating interactions and exhibits first-order phase transitions, so that it is difficult to treat this system using conventional serial-temperature Monte-Carlo methods. Thus we adopt the replica-exchange Monte-Carlo method~\cite{Hukushima96}, in which one simulates $N_{\rm R}$ replicas at different temperatures covering not only a low-temperature regime of interest but also a higher-temperature regime above it, and allows configurational exchange between the replicas. The inclusion of high-temperature configurations enables the lower-temperature systems to access a broad phase space and to avoid being trapped in local energy minima. We perform the simulations for temperature range 0.5$<$$k_{\rm B}T$ (meV)$<$7 with 200 temperature meshes. Both spins and oxygen positions ($\delta_{i,j}$) are updated and relaxed in the simulation. To achieve efficient updates of the oxygen displacements, we adopt a window for Monte-Carlo sampling of $\delta_{i,j}$, $-W<\delta_{i,j}<W$ with $W$=0.5, within which we generate random numbers for the $\delta_{i,j}$ sampling. We carry out each configurational exchange after every 400 standard Monte-Carlo steps. Typically, we perform 1000 exchanges after the sufficient thermalization steps for systems with $N$=48$\times$48$\times$6 sites along the $x$, $y$ and $z$ axes under the periodic boundary condition.

We identify transition points and magnetic structures by calculating temperature profiles of the specific heat $C_{\rm s}(T)$ for the spin system and the $\gamma$-axis component of the total spin-helicity vector $h_\gamma(T)$ with $\gamma$=$a$, $b$ and $c$. They are respectively calculated by
\begin{eqnarray}
C_{\rm s}(T)&=&\frac{1}{N} 
\partial \langle \mathcal{H}-\mathcal{H}_K\rangle / \partial (k_{\rm B}T),\\
h_\gamma(T)&=&\frac{1}{2N} \langle |\sum_{i} 
(\bm S_i \times \bm S_{i+\hat x} 
+\bm S_i \times \bm S_{i+\hat y})_\gamma |\rangle/S^2. \nonumber \\
\end{eqnarray}
Here the brackets denote the thermal average. 

Using the point-charge model, we calculate the electric polarization $\bm P_{\rm S}$=($\tilde P_a$, $\tilde P_b$, $\tilde P_c$) due to the oxygen displacements $\delta_{i,j}$ induced by the ($\bm S \cdot \bm S$)-type magnetostriction. Considering the staggered arrangement of the local axes $\bm n_{i,j}$ on the the zigzag Mn-O chain, we calculate $\tilde P_\gamma$ ($\gamma$=$a$, $b$, $c$) by 
\begin{equation}
\tilde P_\gamma=-\frac{\Pi_\gamma}{N} \langle |
\sum_{i}[(-1)^{i_x+i_y+m} 
\delta_{i,i+\hat x} + (-1)^{i_x+i_y+n} \delta_{i,i+\hat y}] |\rangle ,
\end{equation}
where ($m$, $n$)=(0, 0) for $\gamma$=$a$, ($m$, $n$)=(1, 0) for $\gamma$=$b$, and ($m$, $n$)=($i_z$+1, $i_z$+1) for $\gamma$=$c$. Here the constant $\Pi_\gamma$ is calculated to be 4.6$\times$10$^5$ $\mu C$/$m^2$ for $\gamma$=$a$ and $b$, and 4.7$\times$10$^5$ $\mu C$/$m^2$ for $\gamma$=$c$ from the lattice parameters. 

In order to confirm the magnetic structures, we also calculate the spin and spin-helicity correlation functions in the momentum space, $\hat{S}_{\gamma}(\bm k,T)$ and $\hat{H}_{\gamma}(\bm k,T)$ for $\gamma$=$a$, $b$, and $c$. They are calculated by
\begin{equation}
\hat{S}_{\gamma}(\bm k,T)=\frac{1}{N^2} 
\sum_{i,j} \langle S_{\gamma i}S_{\gamma j}\rangle
e^{i\bm k \cdot (\bm r_i-\bm r_j)},
\label{eq:scrrf}
\end{equation}
\begin{equation}
\hat{H}_{\gamma}(\bm k,T)=\frac{1}{N^2} 
\sum_{i,j} \langle h_{\gamma i}h_{\gamma j}\rangle
e^{i\bm k \cdot (\bm r_i-\bm r_j)}.
\label{eq:hcrrf}
\end{equation}
Here $h_{\gamma i}$ is the $\gamma$ component of the local spin-helicity 
vector $\bm h_i=(h_{a i}, h_{b i}, h_{c i})$, which is defined as
\begin{equation}
\bm h_i=\frac{1}{2}
(\bm S_i \times \bm S_{i+\hat x} 
+ \bm S_i \times \bm S_{i+\hat y})/S^2.
\label{eq:hlcty2}
\end{equation}
In the following, we write these correlation functions simply as 
$\hat{S}_{\gamma}(\bm k)$ and $\hat{H}_{\gamma}(\bm k)$ by omitting $T$.

On the other hand, we study ground-state properties of the Hamiltonian (\ref{eq:model1}) by numerically solving the Landau-Lifshitz-Gilbert equation;
\begin{equation}
\frac{\partial \bm S_i}{\partial t}=-\bm S_i \times \bm H^{\rm eff}_i
+ \frac{\alpha_{\rm G}}{S} \bm S_i \times \frac{\partial \bm S_i}{\partial t}.
\label{eq:LLGEQ}
\end{equation} 
The effective local magnetic fields $\bm H^{\rm eff}_i$ acting on the $i$th Mn spin $\bm S_i$ are derived from the spin-derivative of the Hamiltonian $\mathcal{H}$ as
\begin{equation}
\bm H^{\rm eff}_i = - \partial \mathcal{H} / \partial \bm S_i.
\label{eq:EFFMF}
\end{equation}
Here $\alpha_{\rm G}$ is the dimensionless Gilbert-damping coefficient introduced phenomenologically. For the value of $\alpha_{\rm G}$, we take a rather small value of $\alpha_{\rm G}=0.01$ to achieve a slow relaxation towards a real ground-state spin structure with a minimum energy. We solve this equation using the fourth-order Runge-Kutta method after the linearlization. For the convergence, we use thermally relaxed spin configurations obtained in the Monte-Carlo simulations at low temperatures as initial states.

\section{Results}
\label{Sec:Results}
\subsection{Phase Diagram}
\label{Subsec:PhDgm}
\begin{figure}[tdp]
\includegraphics[scale=1.0]{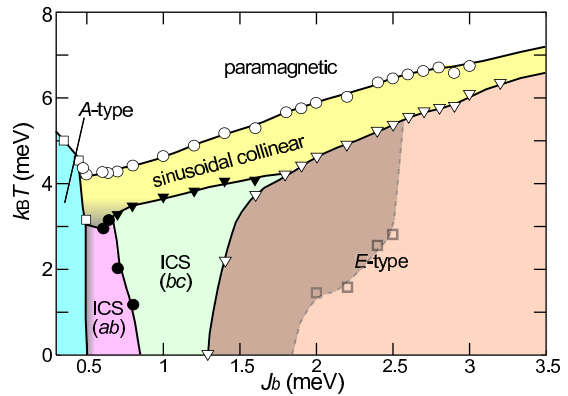}
\caption{(Color online) Theoretical phase diagram of $R$MnO$_3$ in the plane of temperature and $J_b$. Here ICS denotes the incommensurate spiral phase. In the shaded area, incommensurate spin states can coexist with the $E$-type state (see text).}
\label{Fig05}
\end{figure}
In Fig.~\ref{Fig05}, we display theoretically obtained phase diagram in the plane of temperature $k_{\rm B}T$ and antiferromagnetic exchange along the $b$ axis $J_b$~\cite{Mochizuki10b}, which successfully reproduces the experimental phase diagram in Figs.~\ref{Fig02}(a) and (b). At low temperatures, the $A$-type, $ab$-plane spiral, $bc$-plane spiral, and $E$-type phases successively emerge as $J_b$ increases. Here the magnetic structure is commensurate with $q_b$=0.5$\pi$ in the $E$-type phase, whereas it is incommensurate in the $ab$ and $bc$-plane spiral phases. The sinusoidal collinear state is also incommensurate even above the $E$-type phase, and the spin-phonon coupling is a source of the incommensurate-commensurate transition with lowering temperature. In the shaded area, although the $E$-type state has the lowest energy, incommensurate spin states have deep energy minimum, and can coexist with the $E$-type state as will be discussed in Sec.IV-D. 

Note that the experimental phase diagram of the solid-solution systems, i.e., Eu$_{1-x}$Y$_x$MnO$_3$ and Y$_{1-y}$Lu$_y$MnO$_3$ in Fig.~\ref{Fig02}(b) shows that on the verge of the phase boundary between the $bc$-plane spiral and the $E$-type phases, the transition temperature is strongly suppressed, and a V-shaped bicritical point appears. This can be attributed to the randomness effect inherent to the solid-solution systems, which suppresses the transition temperature of the first-order phase transition. Contrastingly, in the experimental phase diagram of the compounds $R$MnO$_3$ with almost no randomness effect [Fig.~\ref{Fig02}(a)], only a very small dip appears. 
The randomness-induced suppression of the first-order transition temperature at the bicritical point has been established well by the early experimental and theoretical studies on the solid-solution Mn perovskites~\cite{Tokura06,Tomioka04}. On the other hand, in the theoretical phase diagram in Fig.~\ref{Fig05}, which is obtained without considering the randomness effect, shows a nearly straight phase boundary with no anomaly. 

\subsection{$E$-type Spin Phase}
\label{Subsec:EtypePh}
\begin{figure}[tdp]
\includegraphics[scale=1.0]{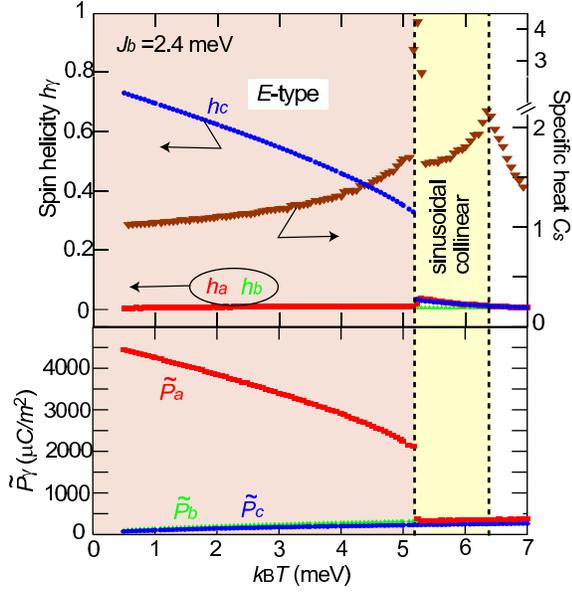}
\caption{(Color online) Temperature profiles of specific heat $C_{\rm s}(T)$, spin helicity $h_\gamma(T)$, and polarizations due to the ($\bm S \cdot \bm S$)-type magnetostriction $\tilde P_\gamma(T)$ for $J_b$=2.4 meV.}
\label{Fig06}
\end{figure}
\begin{figure}[tdp]
\includegraphics[scale=1.0]{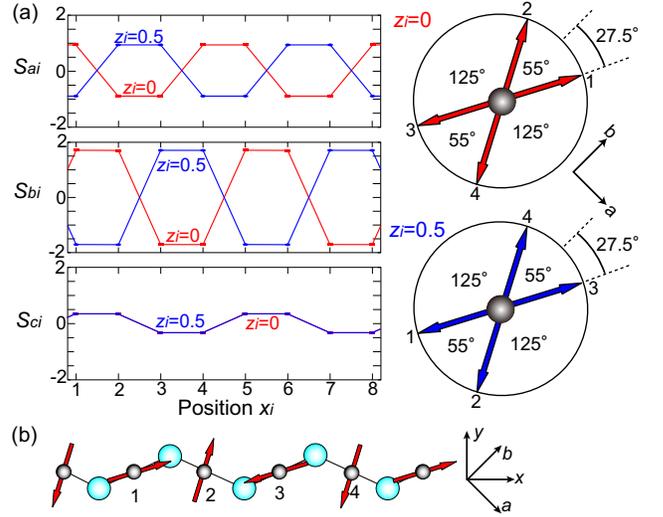}
\caption{(Color online) (a) Real-space spin configuration of the $E$-type phase  for two kinds of $ab$ planes, $z_i$=0 and 0.5. (b) Spin alignment on the zigzag Mn-O chain along the $x$ axis.}
\label{Fig07}
\end{figure}
First we discuss the $E$-type spin phase. In Fig.~\ref{Fig06}, we display calculated temperature profiles of the specific heat $C_{\rm s}(T)$, the spin helicity $h_\gamma(T)$, and the polarization due to the ($\bm S \cdot \bm S$)-type magnetostriction $\tilde P_\gamma(T)$ for $J_b$=2.4 meV. The system exhibits two phase transitions and three magnetic phases emerging successively with lowering temperature, i.e., the paramagnetic, the sinusoidal collinear, and the $E$-type phases~\cite{Kimura06}. The magnetic structure is incommensurate ($q_b$=0.458$\pi$) in the sinusoidal collinear phase, while it is commensurate ($q_b$=0.5$\pi$) in the $E$-type phase.

Interestingly we find a finite $c$-axis component $h_c(T)$ of the spin helicity in the $E$-type phase, indicating that its spin structure is not collinear in reality, but its up-up-down-down structure is subject to a cycloidal deformation within the $ab$ plane. This is in contrast to what has been believed so far. Figure~\ref{Fig07} depicts the real-space spin configuration of the $E$-type order calculated at $T$=0, which indeed shows an elliptically deformed $ab$-plane cycloid. 

\begin{figure}[tdp]
\includegraphics[scale=1.0]{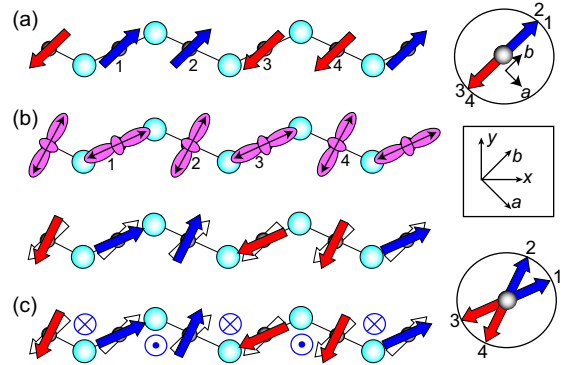}
\caption{(Color online) (a) Pure collinear up-up-down-down spin structure with Mn spins parallel to the $b$ axis. (b) Because of the staggered $d_{3\xi^2-r^2}$/$d_{3\eta^2-r^2}$ orbitals, the local easy magnetization axes are alternately arranged on the zigzag Mn-O chain (upper figure), which can cause deviation of each Mn-spin direction from the $b$ axis towards the local easy axis, resulting in the cycloidal deformation (lower figure). (c) On the zigzag Mn-O chain, the $c$-axis components of the Dzyaloshinskii-Moriya vectors are arranged in a staggered way, with which the Mn spins cant and rotate in the $ab$ plane. Here $\odot$ ($\otimes$) denotes the positive (negative) $c$-axis component of the Dzyaloshinskii-Moriya vector. Note that both the single-ion anisotropy $\mathcal{H}_{\rm sia}^E$ with alternate easy magnetization axes and the Dzyaloshinskii-Moriya interaction $\mathcal{H}_{\rm DM}^{ab}$ give the equivalent spin canting or rotation as can been seen in lower figure of (b) and (c).}
\label{Fig08}
\end{figure}
\begin{figure}[tdp]
\includegraphics[scale=1.0]{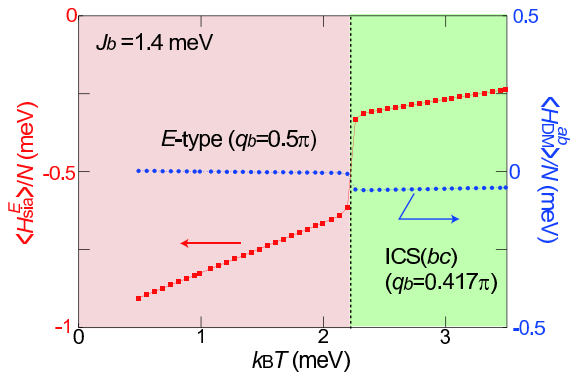}
\caption{(Color online) Temperature dependence of expectation values for $\mathcal{H}_{\rm DM}^{ab}$ and $\mathcal{H}_{\rm sia}^E$ for $J_b$=1.4 meV. At the transition to the $E$-type phase with lowering temperature, the value for $\mathcal{H}_{\rm sia}^E$ abruptly decreases with a large jump, while that for $\mathcal{H}_{\rm DM}$ slightly increases, indicating that the term ${H}_{\rm sia}^E$ is a source of the cycloidal deformation.}
\label{Fig09}
\end{figure}
There are two possible origins for this cycloidal deformation. One is the single-ion anisotropy $\mathcal{H}_{\rm sia}^E$, and the other is the Dzyaloshinskii-Moriya interaction $\mathcal{H}_{\rm DM}$. Let us consider a pure up-up-down-down spin structure shown in Fig.~\ref{Fig08}(a) in which the Mn spins are collinearly aligned parallel to the $b$ axis, and examine what happens if we switch on $\mathcal{H}_{\rm sia}^E$ or $\mathcal{H}_{\rm DM}$. The factor $(-1)^{i_x+i_y}$ in $\mathcal{H}_{\rm sia}^E$ given in Eq.~(6) implies that the local easy magnetization axes are alternately arranged along the in-plane Mn-O chain because of the staggered orbital ordering. Since the occupied $e_g$ orbital is $d_{3\xi^2-r^2}$ or $d_{3\eta^2-r^2}$ , the easy magnetization axis is along the $\xi_i$ or $\eta_i$ axis as shown in upper figure of Fig.~\ref{Fig08}(b) Therefore, in the presence of $\mathcal{H}_{\rm sia}^E$, direction of each Mn spin deviates from the $b$ axis toward $\xi_i$ or $\eta_i$ axis to form an elliptically deformed $ab$-plane cycloid as shown in lower figure of Fig.~\ref{Fig08}(b). On the other hand, in the presence of $\mathcal{H}_{\rm DM}$, the $c$-axis components of the Dzyaloshinskii-Moriya vectors are arranged in a staggered way on the in-plane zigzag Mn-O chains, which can also cause canting of the Mn spins to form a cycloidal rotation as shown in Fig.~\ref{Fig08}(c).

To identify the origin of the cycloidal deformation of the $E$-type spin structure, we calculate temperature-dependence of the expectation values for $\mathcal{H}_{\rm DM}^{ab}$ and $\mathcal{H}_{\rm sia}^E$ at $J_b$=1.4 meV (see Fig.~\ref{Fig09}). Here $\mathcal{H}_{\rm DM}^{ab}$ represents the Dzyaloshinskii-Moriya interaction associated with the vectors $\bm d_{i,j}$ on the in-plane Mn-O bonds, i.e.,
\begin{equation}
\mathcal{H}_{\rm DM}^{ab}=\sum_{i} \bm d_{i,i+\hat x} 
\cdot (\bm S_i \times \bm S_{i+\hat x}) + \sum_{i} \bm d_{i,i+\hat y} 
\cdot (\bm S_i \times \bm S_{i+\hat y}).
\label{eq:INPLANEDM}
\end{equation}
At the transition to the $E$-type phase with lowering temperature, the energy for $\mathcal{H}_{\rm sia}^E$ abruptly decreases with a large jump, while that for $\mathcal{H}_{\rm DM}^{ab}$ increases slightly, indicating the single-ion anisotropy $\mathcal{H}_{\rm sia}^E$ or the alternation of the in-plane easy magnetization axes as its origin. 

The validity of the noncollinearity in the $E$-type phase predicted in the {\it classical} spin model is justified by the large Mn $S$=2 spins. The quantum fluctuation is almost suppressed in $R$MnO$_3$, and this is the reason why our model has successfully described a lot of experimental results for $R$MnO$_3$. More concretely, our model has been established by quantitative reproductions of the phase diagrams~\cite{Mochizuki09a,Mochizuki09b,Mochizuki10b} and the electromagnon optical spectra~\cite{Mochizuki10a}. These facts strongly support robustness of the predicted noncollinear $E$-type order. Furthermore, the predicted extent of the deformation sensitively depends on the strength of the single-ion anisotropy or the parameter value of $E$. Our choice of the parameter $E$=0.25 meV has quantitatively reproduced the area of the sinusoidal collinear phase in the phase diagrams~\cite{Mochizuki09a,Mochizuki09b,Mochizuki10b}, the threshold magnetic field of the field-induced $\bm P$ reorientation~\cite{Mochizuki10c}, the ellipticity of the cycloidal spin structures~\cite{Mochizuki09b}, and the electromagnon spectra~\cite{Mochizuki10a} as experimentally observed, all of which are also sensitive to the value of $E$. These facts guarantee the validity of our parameter choice and, hence, the predicted degree of the noncollinear deformation.

There are already several neutron-scattering studies for $R$MnO$_3$ with $R$=Y, Ho, and Er~\cite{Munoz01,Brinks01,Munoz02,YeF07}, but no apparent noncollinear deformation has been observed. One might think that the noncollinear deformation shown in Fig.~\ref{Fig07} is large enough to be detected experimentally. However it should be mentioned that most of the previous experiments failed to observe the real $E$-type phase. Indeed they reported an incommensurate wave numbers contradicting obviously to the $E$-type state. Thus far, only one experiment by Munoz $et$ $al.$ probably measured the real $E$-type phase with commensurate $q_b$=0.5$\pi$ in HoMnO$_3$~\cite{Munoz01}, but it was performed for powder samples and without any electrical poling procedures. In fact, comparison between calculated Rietveld pattern for the predicted noncollinear $E$-type state and that for the pure collinear $E$-type state in the case of powder sample revealed that only slight differences appear in the intensity of peaks at (0 0.5 1) and (1 0.5 1) with $Pbnm$ setting. The differences are $\sim$10$\%$ for the former peak, while $\sim$12$\%$ for the latter peak~\cite{ArimaP}. In turn, Munoz $et$ $al.$ compared their observed and calculated patterns, which also shows approximately 10$\%$ errors for both peaks (see Fig.11(c) of Ref.~\cite{Munoz01}). These errors suggest that accuracy of their measurement is not enough, or their analysis assuming the {\it collinear} $E$-type state is not appropriate. Now we would like to suggest that there is a possibility to achieve better agreement between the observed and calculated patterns if they perform the analysis assuming the predicted noncollinear $E$-type state. The expected signal of the noncollinearity is very small, and in order to detect such a small difference, a careful sample synthesis, measurement, and analysis are required. Quite recently, Ishiwata and his coworkers have succeeded in synthesizing the perovskite YMnO$_3$ single crystals~\cite{Ishiwata11}, but their sizes ($\sim$0.5 mm) are not large enough for a neutron-scattering experiment.

\begin{figure}[tdp]
\includegraphics[scale=1.0]{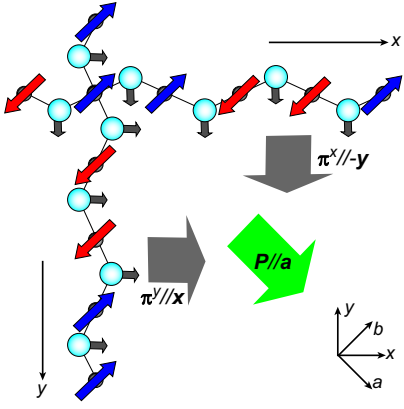}
\caption{(Color online) In the spirally deformed up-up-down-down structure, both ($\bm S \cdot \bm S$)-type and ($\bm S \times \bm S$)-type mechanisms contribute to the ferroelectric $\bm P$. Concerning the former mechanism, alternate large and small spin turn angles cause a uniform shift of the O ions to strengthen and weaken the ferromagnetic exchanges through decreasing and increasing the Mn-O-Mn bond angle, respectively. Shifts of the O ions due to the ($\bm S \cdot \bm S$)-type magnetostriction are shown by small (gray) arrows. Large (red and blue) arrows represent the dominant $b$-axis components of the Mn spins.}
\label{Fig10}
\end{figure}
In the cycloidally deformed $E$-type spin order, both ($\bm S \cdot \bm S$)-type and ($\bm S \times \bm S$)-type mechanisms contribute to the ferroelectric $\bm P$. Concerning the former mechanism, with alternate large and small spin rotation angles or with dominant up-up-down-down spin $b$-axis components, the O ions between nearly (anti)parallel Mn-spin pairs shift negatively (positively) to strengthen (weaken) the ferromagnetic exchanges through increasing (decreasing) the Mn-O-Mn bond angle, which results in the uniform electric polarization. As shown in Fig.~\ref{Fig10}, the oxygen displacements on the zigzag Mn-O chain along the $x$ ($y$) axis contribute to the uniform electric polarization $\bm \pi^x$ ($\bm \pi^y$) pointing in the $-\bm y$ ($\bm x$) direction, respectively. Consequently, the $E$-type spin order generates the ferroelectric $\bm P_{\rm S}$ parallel to the $a$ axis as a sum of $\bm \pi^x$$\parallel$$-\bm y$ and $\bm \pi^y$$\parallel$$+\bm x$.
We also expect a small but finite ($\bm S \times \bm S$) contribution 
$P_{\rm AS}$ due to the cycloidal deformation.

\begin{figure}[tdp]
\includegraphics[scale=1.0]{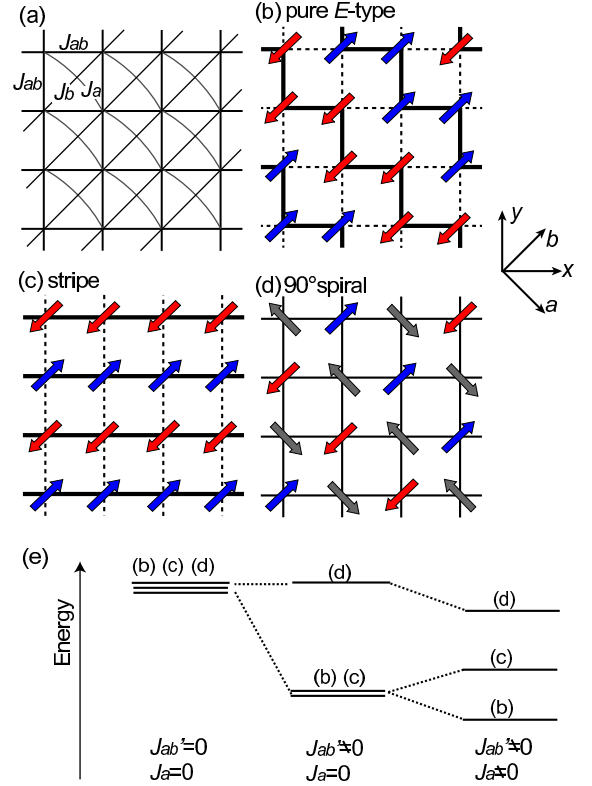}
\caption{(Color online) (a) Spin exchanges of the classical Heisenberg model given by Eq.~(\ref{eq:CHM}). In the limit of large antiferromagnetic $J_b$, three kinds of magnetic states are degenerate, i.e., (b) (pure) $E$-type, (c) stripe, and (d) 90$^{\circ}$ spiral states. The nearest neighbor bonds with energy gain (cost) of $-|J_{ab}|S^2$ ($+|J_{ab}|S^2$) are shown by thick (dotted) lines. The $E$-type state becomes energetically stabilized by bond alternation shown in (b), while the stripe state by that shown in (c). (e) Energy diagram of these three states (see text).}
\label{Fig11}
\end{figure}
Now we discuss roles of the Peierls-type spin-phonon coupling and the weak ferromagnetic coupling $J_a$ along the $a$ axis for realization of the $E$-type spin order. The $E$-type order appears when the antiferromagnetic coupling $J_b$ along the $b$ axis is sufficiently strong. In order to specify the origin of the $E$-type order, we consider the following two-dimensional classical Heisenberg model,
\begin{eqnarray}
\mathcal{H}&=&\sum_{i}
 (J_{ii+\hat{\bm x}}\bm S_i \cdot \bm S_{i+\hat{\bm x}}
+ J_{ii+\hat{\bm y}}\bm S_i \cdot \bm S_{i+\hat{\bm y}}) \nonumber \\
&+&
 J_b\sum_{i} \bm S_i \cdot \bm S_{i+\hat{\bm b}}
+J_a\sum_{i} \bm S_i \cdot \bm S_{i+\hat{\bm a}},
\label{eq:CHM}
\end{eqnarray}
with ferromagnetic $J_{ij}=J_{ab}+J_{ab}'\delta_{i,j}$$<$0, antiferromagnetic $J_b$$>$0, and weakly ferromagnetic $J_a$$<$0 [see also Fig.~\ref{Fig11}(a)]. This is a simplified model to understand the $E$-type order where the single-ion anisotropy and the Dzyaloshinskii-Moriya interaction are removed from the Hamiltonian~(\ref{eq:model1}). When $J_{ab}'$=0 and $J_a$=0, three kinds of magnetic states are degenerate in the limit of strong $J_b$, i.e., the (pure) $E$-type, stripe, and 90$^{\circ}$ spiral states, which are shown in Figs.~\ref{Fig11}(b)-(d), respectively. In all these three states, the spins align antiferromagnetically along the $b$ axis. In the $E$-type and the stripe states, there are energy gains (costs) of $-|J_{ab}|S^2$ ($+|J_{ab}|S^2$) associated with the nearest neighbor ferromagnetic exchanges $J_{ab}$ on the bonds connecting the parallel (antiparallel) spin pairs as indicated by thick (dotted) lines. These gains and costs are perfectly canceled out in total for both cases. On the other hand, there is neither gain nor cost in energy associated with $J_{ab}$ in the 90$^{\circ}$ spiral state because $\bm S_i \cdot \bm S_i$ is always zero in this state. Namely when only the spin-exchange interactions $J_{ab}$ and $J_b$ are considered, the energies of these three states are all identical resulting in their degeneracy. Then if we incorporate the spin-lattice coupling by taking finite $J_{ab}'$, a simultaneous bond alternation sets in to lift the degeneracy by modulating the nearest neighbor ferromagnetic exchanges. The $E$-type and the stripe states become lower in energy by the bond alternations shown in Figs.~\ref{Fig11}(b) and (c), respectively. Here the ferromagnetic exchanges are strengthened (weakened) on the thick (dotted) bonds. In contrast, the energy of the 90$^{\circ}$ spiral state does not change by bond alternations.

The remaining degeneracy of the $E$-type and the stripe states is eventually lifted by the weak ferromagnetic interaction $J_a$. Namely the $E$-type state with ferromagnetically aligned spins along the $a$ axis becomes stabilized by $J_a$, while the stripe state with the staggered spin alignment along the $a$ axis does not. As we have discussed, all of the three terms in the Hamiltonian~(\ref{eq:CHM}) are indispensable for the $E$-type order, and other terms in the original Hamiltonian~(\ref{eq:model1}), i.e., the single-ion anisotropy and the Dzyaloshinskii-Moriya interaction, rather favor the noncollinear spin alignment. Therefore we conclude that the $E$-type order in $R$MnO$_3$ is a consequence of the three ingredients, i.e., the large next neighbor antiferromagnetic coupling $J_b$, the Peierls-type spin-lattice coupling $J_{ab}'\delta_{i,j}$, and the weak ferromagnetic coupling $J_a$.

\begin{figure}[tdp]
\includegraphics[scale=1.0]{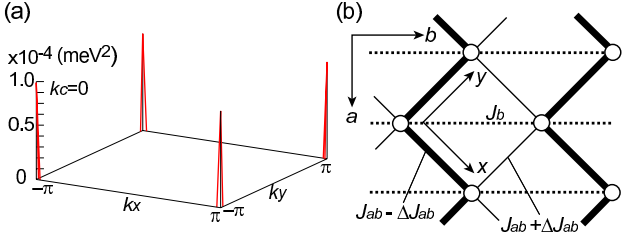}
\caption{(Color online) (a) Calculated correlation function $\hat{J}_{\gamma\gamma'}(\bm k,T)$ given by Eq.~(\ref{eq:Jcrrf}) for the $E$-type state at $J_b$=2.4 meV and $k_{\rm B}T$=0.5 meV, which shows sharp peaks at $\bm k$=($\pm \pi$, $\pm \pi$, 0). Magnitudes of these peaks are all 1.0$\times$10$^{-4}$ meV$^2$. (b) Modulations of the nearest-neighbor ferromagnetic exchanges. The ferromagnetic coupling is strengthened (weakened) by $\Delta J_{ab}$$\sim$0.01 meV on the thick (thin) bonds.}
\label{Fig12}
\end{figure}
To confirm that the bond alternation shown in Fig.~\ref{Fig11}(b) is actually realized in the $E$-type phase, we calculate the following correlation functions at $J_b$=2.4 meV and $k_{\rm B}T$=0.5 meV,
\begin{equation}
\hat{J}_{\gamma\gamma'}(\bm k,T)=\frac{1}{N^2} 
\sum_{i,j} \langle \Delta J_{i,i+\gamma} \Delta J_{j,j+\gamma'}\rangle
e^{i\bm k \cdot (\bm r_i-\bm r_j)},
\label{eq:Jcrrf}
\end{equation}
for ($\gamma$, $\gamma'$)=($x$, $x$), ($y$, $y$) and ($x$, $y$). Here $\Delta J_{i,j}$=$J_{ab}'\delta_{i,j}$. We find that all these correlation functions are identical. They have sharp peaks at $\bm k$=($\pm \pi$, $\pm \pi$, 0), and their magnitudes are all equal to 1.0$\times$10$^{-4}$ meV$^2$ as shown in Fig.~\ref{Fig12}(a). This means that the bond alternation with $J_{ab} \pm \Delta J_{ab}$ is indeed realized where $\Delta J_{ab} \sim \sqrt{\hat{J}_{\gamma\gamma'}(\pi, \pi, 0)}$ is approximately 0.01 meV, i.e., 1.25 $\%$ of the original value $|J_{ab}|$=0.8 meV [see Fig.~\ref{Fig12}(b)].

\subsection{Spiral Spin Phases}
\label{Subsec:SpiralPhs}
\begin{figure*}[tdp]
\includegraphics[scale=1.0]{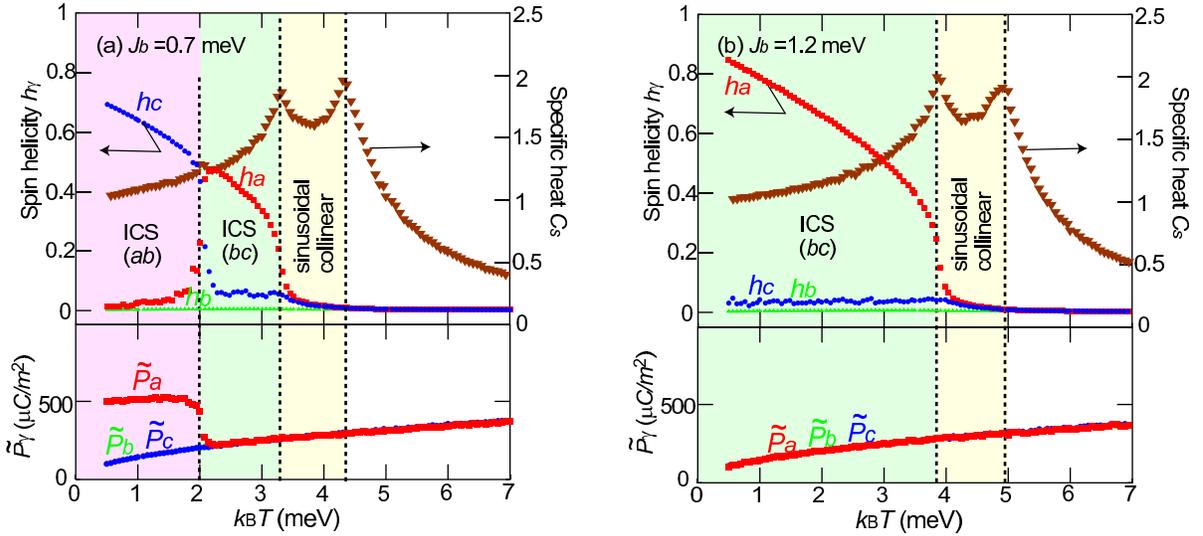}
\caption{(Color online) (a) Temperature profiles of specific heat $C_{\rm s}(T)$, spin helicity $h_\gamma(T)$, and polarizations due to the ($\bm S \cdot \bm S$)-type magnetostriction $\tilde P_\gamma(T)$ for $J_b$=0.7 meV. (b) Those for $J_b$=1.2 meV.}
\label{Fig13}
\end{figure*}
In Figs.~\ref{Fig13}(a) and (b), we show calculated temperature profiles of specific heat $C_{\rm s}(T)$, spin helicity $h_\gamma(T)$, and polarization $\tilde P_\gamma(T)$ due to the ($\bm S \cdot \bm S$)-type magnetostriction for (a) $J_b$=0.7 meV and (b) $J_b$=1.2 meV. For $J_b$=0.7 meV, three successive phase transitions take place with lowering temperature. From the paramagnetic phase, the system first enters into the sinusoidal collinear phase. Subsequently the system enters into the $bc$-plane spiral phase, and finally into the $ab$-plane spiral phase. On the other hand, when $J_b$=1.2 meV, the system exhibits only two phase transitions among three phases, i.e., the paramagnetic, the sinusoidal collinear, and the $bc$-plane spiral phases, whereas the $ab$-plane spiral phase does not appear. 

Interestingly, we see that $\tilde P_a(T)$ in the $ab$-plane spiral phase for $J_b$=0.7 meV is extrapolated to $\sim$500 $\mu C$/$m^2$ at $T$$\rightarrow$0 [lower panel of Fig.~\ref{Fig13}(a)], while to zero in the $bc$-plane spiral phase for $J_b$=1.2 meV [lower panel of Fig.~\ref{Fig13}(b)]. This indicates a finite contribution to the ferroelectric polarization from the ($\bm S \cdot \bm S$)-type magnetostriction in the $ab$-plane spiral phase. Moreover we find that this ($\bm S \cdot \bm S$) contribution can be comparable to or even larger than the ($\bm S \times \bm S$) contribution as discussed later. This is surprising because only the ($\bm S \times \bm S$)-type magnetostriction has been considered as an origin of the ferroelectric order in the spiral spin phase thus far. Contrastingly the ferroelectric order in the $bc$-plane spiral phase is purely of ($\bm S \times \bm S$) origin with no ($\bm S \cdot \bm S$) contribution.

This ($\bm S \cdot \bm S$) contribution can explain puzzling experimental results. It has been known that the amplitude of the ferroelectric $P$ in the $ab$-plane spiral phase is much larger than that in the $bc$-plane spiral phase. For instance, $P$ in the $ab$-plane spiral phase of DyMnO$_3$ under $\bm H$$\parallel$$\bm b$ is 2.5 times larger than $P$ in the $bc$-plane spiral phase at $\bm H$=0~\cite{Kimura05,Kagawa09}. Moreover, in Eu$_{0.6}$Y$_{0.4}$MnO$_3$, the $\bm P$$\parallel$$\bm a$ at $\bm H$=0 is approximately ten times larger than $\bm P$$\parallel$$\bm c$ under $\bm H$$\parallel$$\bm a$~\cite{Yamasaki07b}. Importantly the latter example excludes the possible influence of $f$-electron moments as its origin because of their absence in Eu$^{3+}$ and Y$^{3+}$ ions, which enables us to highlight the roles of spin-phonon coupling on the polarization behavior. These observations have been a puzzle since we expect nearly identical strength of the ($\bm S \times \bm S$)-type magnetostriction in the $ab$-plane and $bc$-plane spiral phases.

\begin{figure}[tdp]
\includegraphics[scale=1.0]{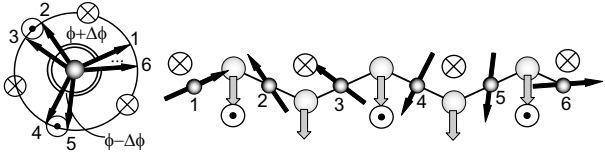}
\caption{Alternation of the spin turn angles in the $ab$-plane spiral state due to the staggered Dzyaloshinskii-Moriya vectors is depicted in an exaggerated manner where $\odot$ ($\otimes$) denotes the positive (negative) $c$-axis component of the vector. Induced shifts of the O ions due to the ($\bm S \cdot \bm S$)-type magnetostriction are shown by gray arrows.}
\label{Fig14}
\end{figure}
The ($\bm S \cdot \bm S$) contribution in the $ab$-plane spiral phase can be understood by a combined function of the Dzyaloshinskii-Moriya interaction and the symmetric magnetostriction (see also Fig.~\ref{Fig14}). On the in-plane zigzag Mn-O chains, the $c$-axis components of the Dzyaloshinskii-Moriya vectors are arranged in the staggered way. Under this circumstance, the spin rotation angles in the $ab$-plane spiral phase become subject to an alternate modulation. Then the O ions between two spins with a smaller angle of $\phi$$-$$\Delta\phi$ (a larger angle of $\phi$$+$$\Delta\phi$) shift negatively (positively) to strengthen (weaken) the ferromagnetic exchange through increasing (decreasing) the Mn-O-Mn bond angle. These shifts generate a uniform component resulting in the ferroelectric polarization. In fact, the spin rotation angles in the $bc$-plane spiral phase are also subject to the alternate modulation because of the staggered $a$-axis components of the Dzyaloshinskii-Moriya vectors. However the induced O shifts are opposite between neighboring $ab$ planes, which results in their perfect cancellation.

The ($\bm S \cdot \bm S$) mechanism in the $ab$-plane spiral phase is triggered by the alternate spin-angle modulation due to the staggered Dzyaloshinskii-Moriya vectors, and thus is a higher-order effect of the Dzyaloshinskii-Moriya interaction. One might think that this ($\bm S \cdot \bm S$) contribution $P_{\rm S}$ cannot be larger than the ($\bm S \times \bm S$) contribution $P_{\rm AS}$ because the latter is a direct consequence of the Dzyaloshinskii-Moriya coupling. But this is not correct, and the $P_{\rm S}$ can be much larger than the $P_{\rm AS}$ in reality. This is because the inherent ``large" staggered Dzyaloshinskii-Moriya vectors cause the ``large" alternate spin-angle modulation, which results in the generation of large $P_{\rm AS}$ through the symmetric ($\bm S \cdot \bm S$)-type magnetostriction. On the other hand, the spiral spin order causes weak ``ferri-components" of the Dzyaloshinskii-Moriya vectors via the antisymmetric ($\bm S \times \bm S$)-type magnetostriction through inducing uniform oxygen shifts or uniform deformations of the electron distribution. However, these ($\bm S \times \bm S$)-induced ferri-components are so small that $P_{\rm AS}$ tends to be small.

Recently Malashevich and Vanderbilt found in their full first-principles calculation that the ferroelectricity in TbMnO$_3$ ($bc$-plane spiral) cannot be explained by the simple ($\bm S \times \bm S$) mechanism only~\cite{Malash08,Malash09}, and they suggested the presence of other coupling. However we should note that the ($\bm S \cdot \bm S$) magnetostriction proposed in the present paper cannot be ``other coupling" suggested by them because the ($\bm S \cdot \bm S$) mechanism works only in the $ab$-plane spiral phase but not in the $bc$-plane spiral phase. The ferroelectricity in TbMnO$_3$ still contains some puzzles in its mechanism, which should be uncovered in the future study.

\begin{figure}[tdp]
\includegraphics[scale=1.0]{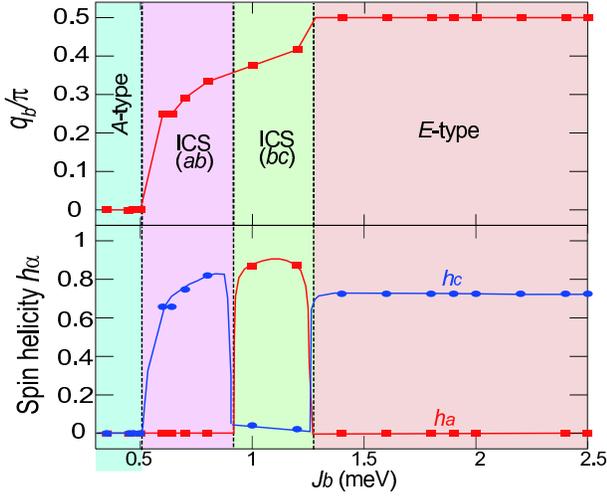}
\caption{(Color online) Calculated $J_b$ dependence of the magnetic wave number $q_b$ and the $\gamma$-axis components of the spin helicity $h_\gamma$ ($\gamma$=$a$, $b$, $c$) at $T$$\rightarrow$0.}
\label{Fig15}
\end{figure}
\begin{figure}[tdp]
\includegraphics[scale=1.0]{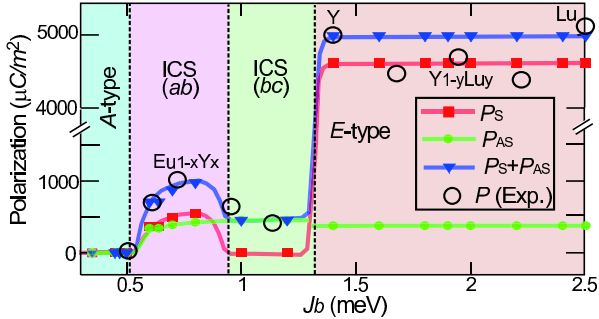}
\caption{(Color online) $J_b$ dependence of polarizations at $T$$\rightarrow$0, i.e., ($\bm S \cdot \bm S$) contribution $P_{\rm S}$, ($\bm S \times \bm S$) contribution $P_{\rm AS}$, and experimentally measured $P$ in Eu$_{1-x}$Y$_x$MnO$_3$ and Y$_{1-y}$Lu$_y$MnO$_3$~\cite{Ishiwata10}. The summation $P_{\rm S}$+$P_{\rm AS}$ reproduces the experimental $P$ well.}
\label{Fig16}
\end{figure}
We also calculate the $J_b$-dependence of the ($\bm S \times \bm S$) contribution $P_{\rm AS}$ at $T$$\rightarrow$0 within the spin-current model where $P_{\rm AS}$ is proportional to the spin helicity $\bm h=\sum_{<i,j>}\bm S_i \times \bm S_j$. Here we use the temperature profile of the spin helicity $h_\gamma(T)$ in Eq.~(17), which is obtained by the Monte-Carlo simulation. Note that the $P_{\rm AS}$ is not calculated from the oxygen positions $\delta_{i,j}$ in contrast to the ($\bm S \cdot \bm S$) contribution $P_{\rm S}$. This is because there are two contributions to $P_{\rm AS}$, i.e, the electronic and the lattice-mediated contributions as pointed out by Malashevich and Vanderbilt~\cite{Malash08,Malash09}. Since the electronic contribution (contribution from the deformed electron-clouds around atoms) cannot be evaluated in our spin model, we calculate the value of $P_{\rm AS}$ at $T$$\rightarrow$0 from $h_\gamma$($T$$\rightarrow$0) shown in Fig.~\ref{Fig15} using the proportional relation between them. The proportionality factor can be evaluated by comparing the calculated value of $h_a$($T$$\rightarrow$0) and the experimental value of $P$ in the $bc$-plane spiral phase because the observed $P$ in the $bc$-plane spiral phase is purely of ($\bm S \times \bm S$) origin. The calculated value of $h_a$($T$$\rightarrow$0) is nearly constant in the $bc$-plane spiral phase and is 0.875 at $J_b$=1.2 meV, while the observed $P_{\rm AS}$($T$$\rightarrow$0) is $\sim$455 $\mu C$/$m^2$ for Eu$_{1-x}$Y$_x$MnO$_3$ with $x$=0.75~\cite{Ishiwata10}. These gives the proportionality factor of 520 $\mu C$/$m^2$.

In Fig.~\ref{Fig16}, we plot the calculated $J_b$-dependence of $P_{\rm S}$, $P_{\rm AS}$, and $P_{\rm S}+P_{\rm AS}$. We also plot experimentally measured $P$ for the solid solutions Eu$_{1-x}$Y$_x$MnO$_3$ and Y$_{1-y}$Lu$_y$MnO$_3$ for comparison~\cite{Ishiwata10}, whose $P$ originates purely from the Mn-spin order because of the absence of $f$ moments. Effective $r_R$ and $J_b$ of these solid solutions are evaluated by interpolations. We find that the sum $P_{\rm S}$+$P_{\rm AS}$ reproduces well the experimental $P$. In particular, our calculation gives constant $P$ in the $E$-type phase in agreement with the experiment. It should be emphasized that only the elastic constant $K$ is an unknown parameter in our model, and once we determine its value so as to reproduce the experimental $P$ in the $E$-type phase, the behaviors of $P$ in the spiral phases are reproduced almost perfectly. Moreover it turns out that the ($\bm S \cdot \bm S$) contribution $P_{\rm S}$ can be comparable to or even larger than the ($\bm S \times \bm S$) contribution $P_{\rm AS}$ in the $ab$-plane spiral phase. This explains why the $P$ in the $ab$-plane spiral phase is much larger than that in the $bc$-plane spiral phase. The 2.5 times larger $\bm P$$\parallel$$\bm a$ under $\bm H$$\parallel$$\bm b$ than $\bm P$$\parallel$$\bm c$ at $\bm H$=0 in DyMnO$_3$ is ascribed to this ($\bm S \cdot \bm S$) contribution. We expect that the ($\bm S \cdot \bm S$) contribution is 1.5 times larger than the ($\bm S \times \bm S$) contribution in the $ab$-plane spiral phase of DyMnO$_3$.

\subsection{Coexisting States}
Next we discuss a certain kind of metastable incommensurate spin state and its possible coexistence with the commensurate $E$-type state in the large $J_b$ region.
The spin-phonon coupling or the ($\bm S \cdot \bm S$)-type magnetostriction make the transition between the incommensurate spiral and the $E$-type phases of strong first order. Consequently, some incommensurate states have deep local energy minima even in the $E$-type phase although the energy comparison gives the transition line as indicated by the solid line in Fig.~\ref{Fig05}. This can result in the realization of a metastable incommensurate spin state trapped in a local energy minimum or its coexistence with the $E$-type state. They can easily occur in reality since the system enters into the $E$-type phase necessarily via the incommensurate sinusoidal collinear phase with lowering temperature. 

For 1.4$<$$J_b$ (meV)$<$2.5, since both the $E$-type and the incommensurate states have energy minima, even the replica-exchange Monte-Carlo calculation sometimes fails to reach the real lowest-energy state. Thus we perform calculations starting with certain initial spin configurations. Note that all other calculations are started with random spin configurations. We chose (A) $E$-type spin configuration obtained for $J_b$=2.4 meV and (B) incommensurate spiral states obtained by switching off the spin-lattice coupling or by setting $J_{ab}'$=0 as initial configurations. In both cases, we perform the Monte-Carlo sampling after sufficient thermalization. 

\begin{figure}[tdp]
\includegraphics[scale=1.0]{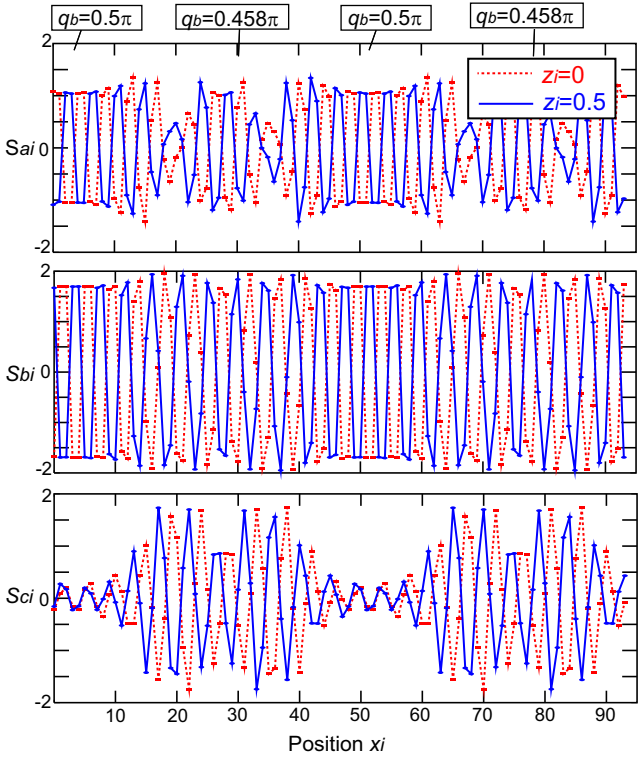}
\caption{(Color online) Real-space spin configuration obtained in the Monte-Carlo simulation starting with an incommensurate spiral spin configuration as the initial state at $J_b$=2.4 meV (see text). The spin $\gamma$-axis components ($\gamma$=$a$, $b$, $c$) are plotted for two kinds of $ab$ plane, i.e., $z_i$=0 and $z_i$=0.5.}
\label{Fig17}
\end{figure}

In Fig.~\ref{Fig17}, we show the real-space spin configuration obtained in the calculation starting with (B). We can see that small incommensurate ($q_b$=0.458$\pi$) spiral regimes exist in the background commensurate ($q_b$=0.5$\pi$) phase. Interestingly we find that these incommensurate regimes emerge periodically to form a stripe structure, indicating possible realization of magnetic discommensulation.

\begin{figure}[tdp]
\includegraphics[scale=1.0]{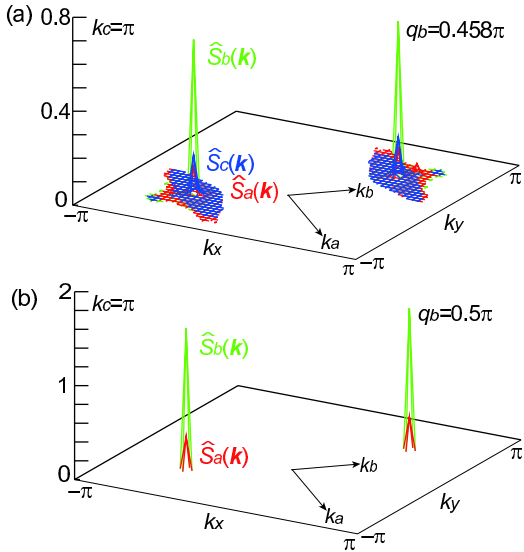}
\caption{(Color online) (a) Calculated spin-correlation functions in the momentum space at $J_b$=2.4 meV for a metastable incommensurate state obtained in the Monte-Carlo simulation starting with the incommensurate spiral state (see text). Their peaks are located at $q_b$=$\pm$0.458$\pi$. (b) Those for the commensurate $E$-type state whose peaks are located at $q_b$=$\pm$0.5$\pi$. Here $\hat S_\gamma(\bm k)$ denotes the correlation function for the spin $\gamma$-axis components given by Eq.~(\ref{eq:scrrf}).}
\label{Fig18}
\end{figure}
In fact, there exists an apparent contradiction in the neutron-diffraction results for $R$MnO$_3$ with small $R$ ions~\cite{Munoz01,Brinks01,Munoz02,YeF07}. Most of the previous experiments reported incommensurate wave numbers $q_b$$\sim$0.43$\pi$ in HoMnO$_3$~\cite{Brinks01}, YMnO$_3$~\cite{Munoz02}, and ErMnO$_3$~\cite{YeF07}. Moreover one of the reports claimed that the magnetic structure in YMnO$_3$ is simultaneously incommensurate and collinear even down to the lowest temperature of 1.7 K~\cite{Munoz02}. On the other hand, one experiment reported a commensurate wave number of $q_b$=0.5$\pi$ in HoMnO$_3$~\cite{Munoz01}. This puzzle can be solved by considering the presence of above metastable incommensurate spin state. We calculate the spin-correlation functions in the momentum space for the above metastable solution. We find that only the spin $b$-axis component has sharp peaks at $q_b$=$\pm$0.458$\pi$, while the other components have very small peaks as shown in Fig.~\ref{Fig18}(a). This seems as if the spin structure were incommensurate {\it collinear}. For comparison we also display the calculated spin correlation function for the commensurate $E$-type state in Fig.~\ref{Fig18}(b) where the spin $b$-axis ($a$-axis) components have large (small) peaks at $q_b$=$\pm$0.5$\pi$. Accordingly the observed incommensurate wave numbers and the claimed incommensurate collinear state in YMnO$_3$ can be attributed to the metastable incommensurate state, while a report of the commensurate $q_b$=0.5$\pi$ in HoMnO$_3$~\cite{Munoz01} can be ascribed to the pure $E$-type state. The metastable incommensurate spin state and its coexistence with the $E$-type state should be seriously considered also when we interpret the experimental results for $R$MnO$_3$ with $R$=Y, Ho, ...,Lu, such as strange electromagnon spectra in the THz optical spectroscopy~\cite{Takahashi10}.

\section{Summary}
In summary, we have theoretically studied the origins and nature of the multiferroic phases and the magnetoelectric coupling in $R$MnO$_3$ by using a realistic spin model including the Peierls-type spin-phonon coupling, which can successfully reproduces the entire phase diagram of $R$MnO$_3$. We have revealed the cooperative contributions of symmetric ($\bm S \cdot \bm S$)-type and antisymmetric ($\bm S \times \bm S$)-type magnetostrictions to the ferroelectricity in the $ab$-plane spiral phase. This large ($\bm S \cdot \bm S$) contribution to ferroelectricity is expected and should be seriously considered also in other spin-spiral-based multiferroics, for which the ($\bm S \times \bm S$)-type magnetostriction has been believed to be a unique origin of the ferroelectric polarization. We have also uncovered the cycloidal spin deformation in the $E$-type phase due to the alternate arrangement of the easy-magnetization axes on the in-plane zigzag MnO chain. We expect that this cycloidal deformation would be detected in a future neutron-scattering experiment. A metastable incommensurate spin state and its coexistence with the $E$-type state have been also found on the verge of the phase boundary between the $E$-type and spiral states. On these basis, a puzzle in the neutron-scattering experiments for $R$=Y, Ho, and Er have been resolved. Our model gives a firm basis for studying and controlling the intriguing cross-correlation phenomena in $R$MnO$_3$. Moreover the crucial roles of the spin-phonon coupling we have demonstrated by taking $R$MnO$_3$ are not specific to this manganite system, but are relevant to all of the multiferroic materials.

\section*{Acknowledgment}
The authors are grateful to Y. Tokura, S. Ishiwata, F. Kagawa, D. Okuyama, and T. Arima for discussions. This work was supported by Grant-in-Aid for Scientific Research (Grants No. 22740214, No. 21244053, No. 17105002, No. 19048015, and No. 19048008), Global-COE Program (``Physical Sciences Frontier") and NAREGI Project from the Ministry of Education, Culture, Sports, Science and Technology of Japan, and by Funding Program for World-Leading Innovative R$\&$D on Science and Technology (FIRST Program) on ``Quantum Science on Strong Correlation".



\begin{thebibliography}{999}
\bibitem{Dzyaloshinskii59}I. E. Dzyaloshinskii, Zh. Eksp. Teor. Fiz. {\bf 33}, 881 (1959) [Sov. Phys. JETP {\bf 10}, 628 (1960)].

\bibitem{Astrov60}D. N. Astrov, Zh. Eksp. Teor. Fiz. {\bf 38}, 984 (1960) [Sov. Phys. JETP {\bf 11}, 708 (1960)].

\bibitem{Freeman75}{\it Magnetoelectric Interaction Phenomena in Crystals}, edited by A.J.Freeman and H. Schmid (Gordon and Breach, London, 1975).

\bibitem{Kimura03a}T. Kimura, T. Goto, H. Shintani, K. Ishizaka, T. Arima, and Y. Tokura, Nature (London) {\bf 426}, 55 (2003).

\bibitem{Schmid94}H. Schmid, Ferroelectrics {\bf 162}, 317 (1994).

\bibitem{Hill00}N. A. Hill, J. Phys. Chem. B {\bf 104}, 6694 (2000).

\bibitem{Fiebig05}M. Fiebig, J. Phys. D: Appl. Phys. {\bf 38}, R123 (2005).

\bibitem{Khomskii06}D. I. Khomskii, J. Magn. Magn. Mater. {\bf 306}, 1 (2006).

\bibitem{Tokura06a}Y. Tokura, Science {\bf 312}, 1481 (2006).

\bibitem{Eerenstein06}W. Eerenstein, N. D. Mathur, and J. F. Scott, Nature (London) {\bf 442}, 759 (2006).

\bibitem{Cheong07}S.-W. Cheong and M. Mostovoy, Nat. Mater. {\bf 6}, 13 (2007).

\bibitem{Tokura07}Y. Tokura, J. Magn. Magn. Mater. {\bf 310}, 1145 (2007).

\bibitem{Pimenov06a}A. Pimenov, A. A. Mukhin, V. Yu. Ivanov, V. D. Travkin, A. M. Balbashov, and A. Loidl, Nat. Phys. {\bf 2}, 97 (2006).

\bibitem{Kida09}N. Kida, Y. Takahashi, J. S. Lee, R. Shimano, Y. Yamasaki, Y. Kaneko, S. Miyahara, N. Furukawa, T. Arima, and Y. Tokura,
J. Opt. Soc. Am. B {\bf 26}, A35 (2009).

\bibitem{Aguilar09}R. ValdesAguilar, M. Mostovoy, A. B. Sushkov, C. L. Zhang, Y. J. Choi, S-W. Cheong, and H. D. Drew, Phys. Rev. Lett. {\bf 102}, 047203 (2009).

\bibitem{Mochizuki10a}M. Mochizuki, N. Furukawa, and N. Nagaosa,
Phys. Rev. Lett. {\bf 104}, 177206 (2010).

\bibitem{Mochizuki10d}M. Mochizuki and N. Nagaosa,
Phys. Rev. Lett. {\bf 105}, 147202 (2010).

\bibitem{Kimura05}T. Kimura, G. Lawes, T. Goto, Y. Tokura, and A. P. Ramirez, Phys. Rev. B {\bf 71}, 224425 (2005).

\bibitem{MTokunaga09}M. Tokunaga, Y. Yamasaki, Y. Onose, M. Mochizuki, N. Furukawa, and Y. Tokura, Phys. Rev. Lett. {\bf 103}, 187202 (2009).

\bibitem{Abe07}N. Abe, K. Taniguchi, S. Ohtani, T. Takenobu, Y. Iwasa, and T. Arima, Phys. Rev. Lett. {\bf 99}, 227206 (2007).

\bibitem{Murakawa08b}H. Murakawa, Y. Onose, F. Kagawa, S. Ishiwata, Y. Kaneko, and Y. Tokura, Phys. Rev. Lett. {\bf 101}, 197207 (2008).

\bibitem{Mochizuki10c}M. Mochizuki and N. Furukawa,
Phys. Rev. Lett. {\bf 105}, 187601 (2010).

\bibitem{Goto04}T. Goto, T. Kimura, G. Lawes, A. P. Ramirez, and Y. Tokura, Phys. Rev. Lett. {\bf 92}, 257201 (2004).

\bibitem{Kagawa09}F. Kagawa, M. Mochizuki, Y. Onose, H. Murakawa, Y. Kaneko, N. Furukawa, and Y. Tokura, Phys. Rev. Lett.{\bf 102}, 057604 (2009).

\bibitem{Schrettle09}F. Schrettle, P. Lunkenheimer, J. Hemberger, V. Yu. Ivanov, A. A. Mukhin, A. M. Balbashov, and A. Loidl, Phys. Rev. Lett. {\bf 102}, 207208 (2009).

\bibitem{Katsura05}H. Katsura, N. Nagaosa, and A. V. Balatsky, Phys. Rev. Lett. {\bf 95}, 057205 (2005).

\bibitem{Sergienko06a}I. A. Sergienko and E. Dagotto, Phys. Rev. B {\bf 73}, 094434 (2006).

\bibitem{Mostovoy06}M. Mostovoy, Phys. Rev. Lett. {\bf 96}, 067601 (2006).

\bibitem{Kenzelmann05}M. Kenzelmann, A. B. Harris, S. Jonas, C. Broholm, J. Schefer, S. B. Kim, C. L. Zhang, S.-W. Cheong, O. P. Vajk, and J. W. Lynn, Phys. Rev. Lett. {\bf 95}, 087206 (2005).

\bibitem{Yamasaki07a}Y. Yamasaki, H. Sagayama, T. Goto, M. Matsuura, K. Hirota, T. Arima, and Y. Tokura, Phys. Rev. Lett. {\bf 98}, 147204 (2007).

\bibitem{Arima06}T. Arima, A. Tokunaga, T. Goto, H. Kimura, Y. Noda, and Y. Tokura, Phys. Rev. Lett. {\bf 96}, 097202 (2006).

\bibitem{Yamasaki08}Y. Yamasaki, H. Sagayama, N. Abe, T. Arima, K. Sasai, M. Matsuura, K. Hirota, D. Okuyama, Y. Noda, and Y. Tokura, Phys. Rev. Lett. {\bf 101}, 097204 (2008).

\bibitem{Mochizuki09a}M. Mochizuki, and N. Furukawa, J. Phys. Soc. Jpn. {\bf 78}, 053704 (2009).

\bibitem{Mochizuki09b}M. Mochizuki, and N. Furukawa, Phys. Rev. B {\bf 80}, 134416 (2009).

\bibitem{Sergienko06b}I. A. Sergienko, C. Sen, and E. Dagotto, Phys. Rev. Lett. {\bf 97}, 227204 (2006).

\bibitem{Picozzi07}S. Picozzi, K. Yamauchi, B. Sanyal, I. A. Sergienko, and E. Dagotto, Phys. Rev. Lett. {\bf 99}, 227201 (2007).

\bibitem{Yamauchi08}K. Yamauchi, F. Freimuth, S. Blugel, and S. Picozzi, Phys. Rev. B {\bf 78}, 014403 (2008).

\bibitem{Ishiwata10}S. Ishiwata, Y. Kaneko, Y. Tokunaga, Y. Taguchi, T. Arima, and Y. Tokura, Phys. Rev. B {\bf 81}, 100411(R) (2010).

\bibitem{Ishiwata11}S. Ishiwata, Y. Tokunaga, Y. Taguchi, and Y. Tokura, J. Am. Chem. Soc. {\bf 133}, 13818 (2011).

\bibitem{Lorenz07}B. Lorenz, Y. Q Wang, and C. W. Chu, Phys. Rev. B {\bf 76}, 104405 (2007).

\bibitem{Pomjakushin09}V. Yu. Pomjakushin, M. Kenzelmann, A. Donni, A. B. Harris, T. Nakajima, S. Mitsuda, M. Tachibana, L. Keller, J. Mesot, H. Kitazawa, and E. Takayama-Muromachi, New J. of Phys. {\bf 11}, 043019 (2009).

\bibitem{Kaplan11}T. A. Kaplan and S. D. Mahanti, Phys. Rev. B {\bf 83}, 174432 (2011).

\bibitem{Goto05}T. Goto, Y. Yamasaki, H. Watanabe, T. Kimura, and Y. Tokura, Phys. Rev. B {\bf 72}, 220403(R) (2005).

\bibitem{Hemberger07}J. Hemberger, F. Schrettle, A. Pimenov, P. Lunkenheimer, V. Y. Ivanov, A. A. Mukhin, A. M. Balbashov, and A. Loidl, Phys. Rev. B {\bf 75}, 035118 (2007).

\bibitem{Yamasaki07b}Y. Yamasaki, S. Miyasaka, T. Goto, H. Sagayama, T. Arima, and Y. Tokura, Phys. Rev. B {\bf 76}, 184418 (2007).

\bibitem{Ivanov06}V. Yu. Ivanov, A. A. Mukhin, V. D. Travkin, A. S. Prokhorov, A. M. Kadomtsev, Yu. F. Popov, G. P. Vorobev, K. I. Kamilov, and A. M. Balbashov, J. Magn. Magn. Mater. {\bf 300}, e130 (2006).

\bibitem{Ivanov06b}V. Yu. Ivanov, A. A. Mukhin, V. D. Travkin, A. S. Prokhorov, Yu. F. Popov, A. M. Kadomtseva, G. P. Vorob\'ev, K. I. Kamilov, A. M. Balbashov, Phys. Status Solidi B {\bf 243}, 107 (2006).

\bibitem{Kimura03b}T. Kimura, S. Ishihara, H. Shintani, T. Arima, K. T. Takahashi, K. Ishizaka, and Y. Tokura, Phys. Rev. B {\bf 68}, 060403(R) (2003).


\bibitem{DongS08}
T. Hotta, M. Moraghebi, A. Feiguin, A. Moreo, S. Yunoki, and E. Dagotto, Phys. Rev. Lett. {\bf 90}, 247203 (2003);
S. Dong, R. Yu, S. Yunoki, J.-M. Liu, and E. Dagotto, Phys. Rev. B {\bf 78}, 155121 (2008).

\bibitem{Kaplan09}T. A. Kaplan, Phys. Rev. B {\bf 80}, 012407 (2009).

\bibitem{Hayden10}L. X. Hayden, T. A. Kaplan, and S. D. Mahanti, Phys. Rev. Lett. {\bf 105}, 047203 (2010).

\bibitem{Furukawa10}N. Furukawa, and M. Mochizuki, J. Phys. Soc. Jpn. {\bf 79}, 033708 (2010).

\bibitem{Mochizuki10b}M. Mochizuki, N. Furukawa, and N. Nagaosa,
Phys. Rev. Lett. {\bf 105}, 037205 (2010).

\bibitem{Alonso00}J. A. Alonso, M. J. Mart\'inez-Lope, M. T. Casais, and M. T. Fern\'andez-D\'iaz, Inorg. Chem. {\bf 39}, 917 (2000).

\bibitem{Dzyaloshinsky58}I. Dzyaloshinsky, J. Phys. Chem. Solids {\bf 4}, 241 (1958).
\bibitem{Moriya60a}T. Moriya, Phys. Rev. Lett. {\bf 4}, 228 (1960).
\bibitem{Moriya60b}T. Moriya, Phys. Rev. {\bf 120}, 91 (1960).

\bibitem{Solovyev96}I. Solovyev, N. Hamada, and K. Terakura, Phys. Rev. Lett. {\bf 76}, 4825 (1996).

\bibitem{Picozzi06}S. Picozzi, K. Yamauchi, G. Bihlmayer, and S. Blugel, Phys. Rev. B {\bf 74}, 094402 (2006).

\bibitem{Hukushima96}K. Hukushima and K. Nemoto, J. Phys. Soc. Jpn. {\bf 65}, 1604 (1996).

\bibitem{Tokura06}Y. Tokura, Reports on Progress in Physics {\bf 69}, 797 (2006).

\bibitem{Tomioka04}Y. Tomioka and Y. Tokura, Phys. Rev. B {\bf 70}, 014432 (2004).

\bibitem{Kimura06}Successive emergence of these three magnetic phases was also observed in other magnetoelectric system, CuFeO$_2$; T. Kimura, J. C. Lashley, and A. P. Ramirez, Phys. Rev. B {\bf 73}, 220401 (2006).

\bibitem{Munoz01}A. Munoz, M. T. Casais, J. A. Alonso, M. J. Mart\'inez-Lope, J. L. Mart\'inez, and M. T. Fern\'andez-D\'iaz, Inorg. Chem. {\bf 40}, 1020 (2001).

\bibitem{Brinks01}H. W. Brinks, J. Rodr\'iguez-Carvajal, H. Fjellvag, A. Kjekshus, and B. C. Hauback, Phys. Rev. B {\bf 63}, 094411 (2001).

\bibitem{Munoz02}A. Munoz, J. A. Alonso, M. T. Casais, M. J. Mart\'inez-Lope, J. L. Mart\'inez, and M. T. Fern\'andez-D\'iaz, J. Phys.: Condens. Matter {\bf 14}, 3285 (2002).

\bibitem{YeF07}F. Ye, B. Lorenz, Q. Huang, Y. Q. Wang, Y. Y. Sun, C. W. Chu, J. A. Fernandez-Baca, Pengcheng Dai, and H. A. Mook, Phys. Rev. B {\bf 76}, 060402(R) (2007).

\bibitem{ArimaP}T. Arima and D. Okuyama, private communication.

\bibitem{Malash08}A. Malashevich and D. Vanderbilt, Phys. Rev. Lett. {\bf 101}, 037210 (2008).

\bibitem{Malash09}A. Malashevich and D. Vanderbilt, Phys. Rev. B {\bf 80}, 224407 (2009).
\bibitem{Takahashi10}Y. Takahashi, S. Ishiwata, S. Miyahara, Y. Kaneko, N. Furukawa, Y. Taguchi, R. Shimano, and Y. Tokura, Phys. Rev. B {\bf 81}, 100413 (2010).

\end{thebibliography}
\end{document}